\documentclass[aps,floats]{revtex4}
\usepackage{amsmath,amssymb}
\usepackage{graphicx,epsfig}
\usepackage[greek,english]{babel}
\usepackage{bbold}

\begin{document}
\bibliographystyle {plain}

\pdfoutput=1
\def\oppropto{\mathop{\propto}} 
\def\opsimeq{\mathop{\simeq}}
\def\opoverderline{\mathop{\overline}}
\def\operarrow{\mathop{\longrightarrow}}
\def\opsim{\mathop{\sim}}

\def\fig#1#2{\includegraphics[height=#1]{#2}}
\def\figx#1#2{\includegraphics[width=#1]{#2}}


\title{ Inference of Markov models from trajectories via Large Deviations at Level 2.5 
\\ with applications to random walks in disordered media } 


\author{ C\'ecile Monthus }
 \affiliation{Institut de Physique Th\'{e}orique, 
Universit\'e Paris Saclay, CNRS, CEA,
91191 Gif-sur-Yvette, France}

\begin{abstract}
The inference of Markov models from data on stochastic dynamical trajectories over the large time-window $T$ is revisited via the Large Deviations at Level 2.5 for the time-empirical density and the time-empirical flows. The goal is to obtain the large deviations properties for the probability distribution of the inferred Markov parameters in order to characterize their possible fluctuations around the true Markov parameters for large $T$. The explicit rate functions are given for several settings, namely discrete-time Markov chains, continuous-time Markov jump processes, and diffusion processes in dimension $d$. Applications to various models of random walks in disordered media are described, where the goal is to infer the quenched disordered variables defining a given disordered sample.

\end{abstract}

\maketitle

\section{ Introduction }

Inference has always played a major role in probability and statistics 
\cite{book_information,book_learning,book_inference}.
However the recent availability of big data in many fields has triggered an enormous increase 
of applications of inference methods.
In particular, the inference of Markov models from data on stochastic dynamical trajectories
has been implemented within various settings \cite{bayesmarkov}
including discrete-time Markov chains \cite{1957,1961,zeitouni},
continuous-time Markov jump processes \cite{evolution,monassonDNAprl,monassonDNApre,monassonsinai2008,monassonsinai2009},
and Langevin dynamics \cite{noise,mapsforce,biomolecule,singlemolecule,bacterial,decision,opticaltweezer,membrane,proteins,diffcoef,primerbayesian,metzler,cherstvy,movies,ronceray2020,underdamped,secondorder,learning,cellcell},
as well as for several active matter models \cite{flocking,activeswimmwe,activematter}.

Besides the development of efficient numerical inference procedures,
it is thus important to characterize theoretically the statistical fluctuations of these inferred parameters of Markov models
with respect to the 'true' values of parameters.
The goal of the present paper is to analyze their large deviations properties
with respect to the time-window size $T$ of the observed dynamical trajectories.
Within the recent progresses in the field of non-equilibrium stochastic processes
(see the reviews with different scopes \cite{derrida-lecture,harris_Schu,searles,harris,mft,sollich_review,lazarescu_companion,lazarescu_generic,jack_review}, the PhD theses \cite{fortelle_thesis,vivien_thesis,chetrite_thesis,wynants_thesis} 
 and the HDR thesis \cite{chetrite_HDR}),
 the theory of large deviations (see the reviews \cite{oono,ellis,review_touchette} and references therein)
 plays the role of the unifying language.
 However the usual classification into three levels, namely Level 1 for empirical observables, 
 Level 2 for the empirical measure, and Level 3 for the empirical process has turned out to be insufficient.
 So the new intermediate 'Level 2.5' 
 concerning the joint distribution of the empirical measure and of the empirical flows
 has emerged as the appropriate level for non-equilibrium steady-states,
 where one can write explicit expressions for the rate functions.
This large deviation analysis at Level 2.5 has been applied to various settings,
including discrete-time Markov chains 
 \cite{fortelle_thesis,fortelle_chain,review_touchette,c_largedevdisorder,c_reset},
continuous-time Markov jump processes 
\cite{fortelle_thesis,fortelle_jump,maes_canonical,maes_onandbeyond,wynants_thesis,chetrite_formal,BFG1,BFG2,chetrite_HDR,c_ring,c_interactions,c_open,barato_periodic,chetrite_periodic,c_reset}
and diffusion processes 
\cite{wynants_thesis,maes_diffusion,chetrite_formal,engel,chetrite_HDR,c_reset,c_lyapunov}.
In the present paper, these large deviations properties at Level 2.5 for the empirical observables of dynamical trajectories
are translated into the large deviations for the inferred parameters of Markov models.
Applications to some models of random walks in disordered media 
 (see the reviews \cite{haus,jpb_review,annphys90,havlin,c_review})
 are described, where the goal is to infer the quenched disordered variables defining a given disordered sample.


The paper is organized as follows.
Section \ref{sec_general} explains the general framework to 
analyze the large deviations properties for the inference of Markov models from dynamical trajectories.
The case of Markov chains in discrete time is discussed in section \ref{sec_chain},
with the application to the random walk on the disordered ring in section \ref{sec_RWRing}.
The case of Markov jump processes in continuous time is presented in section \ref{sec_jump},
with the application to the directed trap model on a disordered ring in section \ref{sec_TrapRing}.
Finally, the case of diffusion processes in dimension $d$ is described in section \ref{sec_diff}
with the application to the diffusion in a disordered potential in section \ref{sec_diffpotential}.
Our conclusions are summarized in section \ref{sec_conclusion}.
Appendix \ref{app_pathintegral} contains more technical details on the path-integral analysis for diffusion processes.


\section{ General principles to infer Markov models from dynamical trajectories  }

\label{sec_general}

Before the analysis of various specific Markov models (discrete or continuous time, discrete or continuous space) in the next sections,
it seems useful in the present section to outline the general principles.

\subsection{ Goals and Notations }

The guideline of the present paper can be summarized in terms of the following unifying notations : 
\\

(i) $M$ is the 'true' Markov model with steady state. Its parameters are unknown and need to be inferred from data.
\\

(ii) $D$ are the data concerning a single dynamical trajectory $x(0 \leq t \leq T)$ of the model $M$ over the long time $T$.
\\

(iii) $E$ are the relevant time-empirical observables : they can be computed as the averages of simple time-local operators
over the trajectory $x(0 \leq t \leq T)$ and they are sufficient to evaluate the probability of the trajectory $x(0 \leq t \leq T) $ for the model $M$. The large deviation properties of these relevant time-empirical observables are explicitly known
for many Markov models $M$ and are called 'Large deviations at Level 2.5', as already mentioned in the introduction.
\\

(iv) $\hat M (E)$ is the best Markov model that can be inferred via the maximum-likelihood method 
from the values $E$ of the relevant time-empirical observables 
computed from the data $D$ concerning the single dynamical trajectory $x(0 \leq t \leq T)$.
The large deviation properties at Level 2.5 of the empirical observables $E$ mentioned in (iii) can be translated into
the large deviation properties of the inferred model $\hat M$ in order to characterize
the fluctuations of inferred model $\hat M $ around the true model $M$ for large $T$.
\\

Let us now describe in more details the various steps of such an analysis.

\subsection{ Identification of the relevant time-empirical observables that determine the trajectories probabilities }

For the 'true' Markov model $M$, 
the first step consists in rewriting the probability of a long dynamical trajectory $x(0 \leq t \leq T)$ 
\begin{eqnarray}
 {\cal P}[x(0 \leq t \leq T)] \opsimeq_{T \to +\infty}   e^{\displaystyle  -T  A_{[M]} \left(E [x(0 \leq t \leq T)] \right) }
\label{ptrajectempi}
\end{eqnarray}
in terms of an intensive action $A_{[M]} \left( E [x(0 \leq t \leq T)] \right) $ that depends on the model parameters $M$,
and that only involves a few relevant time-empirical observables $E [x(0 \leq t \leq T)] $ of the dynamical trajectory $x(0 \leq t \leq T) $.

\subsection{ Number of dynamical trajectories of length $T$ with the same value of the time-empirical observables }

Since all the individual dynamical trajectories $ x(0 \leq t \leq T)  $
 that have the same empirical observables $E=E [x(0 \leq t \leq T)] $
  have the same probability given by Eq. \ref{ptrajectempi},
 one can rewrite the normalization over all possible trajectories 
as a sum over these empirical observables
\begin{eqnarray}
1= \sum_{x(0 \leq t \leq T)}  {\cal P}[x(0 \leq t \leq T)] 
\opsimeq_{T \to +\infty}  
 \sum_{E} \Omega_T(E ) e^{\displaystyle  -T  A_{[M]} \left( E  \right) }
\label{normaempi}
\end{eqnarray}
where the number of dynamical trajectories of length $T$ associated to given values $E $ of these 
empirical observables
\begin{eqnarray}
\Omega_T ( E ) \equiv \sum_{x(0 \leq t \leq T)} \delta \left(E [x(0 \leq t \leq T)] - E \right)
\label{omegaaempi}
\end{eqnarray}
 is expected to grow exponentially with respect to the length $T$ of the trajectories
\begin{eqnarray}
 \Omega_T( E) \opsimeq_{T \to +\infty} C(E) \ e^{\displaystyle T S( E )  }
\label{omegat}
\end{eqnarray}
while the prefactor $C(E)$ denotes the appropriate constitutive constraints for the empirical observables.
The factor $S( E )  = \frac{\ln \Omega_T(E) }{ T }  $ represents the 
Boltzmann intensive entropy of the set of trajectories of length $T$ with given intensive empirical observables $E $.
Let us now recall how it can be evaluated without any actual computation (i.e. one does not need 
to use combinatorial methods to count the appropriate configurations).

The normalization of Eq. \ref{normaempi} becomes for large $T$
\begin{eqnarray}
1  \opsimeq_{T \to +\infty} \sum_{E} C(E) \  e^{\displaystyle  T \left[ S( E  ) - A_{[M]} \left( E \right)   \right] }
\label{normaempit}
\end{eqnarray}
When the empirical variables $E $ take their typical values $E_M^{typ}$ for the model $M$,
the exponential behavior in $T$ of Eq. \ref{normaempit}
should exactly vanish,
i.e. the entropy $S( E_M^{typ}  ) $ should exactly compensate the action $A_{[M]} \left( E_M^{typ}\right)   $ 
\begin{eqnarray}
S( E_M^{typ}  )=    A_{[M]} \left( E_M^{typ}\right) 
\label{compensation}
\end{eqnarray}
To obtain the intensive entropy $S( E  ) $ for any other given value $E  $ of the empirical observables,
one just needs to introduce the modified model $\hat M (E)$ that would make 
the empirical values $E $ typical for this modified model
\begin{eqnarray}
   E = E_{\hat M(E)}^{typ}
\label{modeleeff}
\end{eqnarray}
and to use Eq. \ref{compensation} for this modified model to obtain
\begin{eqnarray}
S( E) = S( E_{\hat M (E)}^{typ}  )=    A_{[{\hat M} (E)]} \left(E_{\hat M (E)}^{typ}  \right) = A_{[{\hat M} (E)]} \left(E \right) 
\label{entropyempi}
\end{eqnarray}


\subsection{ Large deviations for the relevant time-empirical observables $E$}

 Eq \ref{normaempi}
means that, for the model $M$, the probability $ P_T (E) $ of the empirical observables $E$ 
over the set of dynamical trajectories of length $T$,
with the normalization
\begin{eqnarray}
&& 1  = \sum_{ E }  P_T  (E)
\label{normaprobaempi}
\end{eqnarray}
follows the large deviation form
\begin{eqnarray}
&& P_T  (E) \opsimeq_{T \to +\infty}  C(E) \ e^{\displaystyle  - T  I_{[M]} (E)    }
\label{probaaempi}
\end{eqnarray}
where the rate function
\begin{eqnarray}
I_{[M]}  (E)  = A_{[M]} \left( E \right)   - S( E  )  
= A_{[M]}  (E) - A_{[{\hat M}(E)]}  (E) 
\label{rateempi}
\end{eqnarray}
is simply given by the difference between the intensive action $A_{[M]}  (E) $ associated to the true model $M$
and the intensive action $A_{[{\hat M}(E)]}  (E) $ associated to the modified model ${\hat M}(E)$ that would make the empirical value $E$ typical (see Eq. \ref{modeleeff}).
It is positive $I_{[M]}  (E) \geq 0 $ and vanishes only for the typical value $E_M^{typ}$
\begin{eqnarray}
0=I_{[M]}  (E_M^{typ}) 
\label{rateempizerotyp}
\end{eqnarray}
i.e. only when the modified model ${\hat M} (E)$ coincides with the true model $M$.


\subsection{ Large deviations for the inferred model $\hat M$ obtained via the principle of maximum likelihood  }

Now we wish to infer the true model $ M$ from the empirical observables $E$ computed from the data $D$.
The likelihood ${\cal L}_T(  M \vert E ) $ of the model $ M$ given $E$
 is defined as the probability $P_T( E \vert  M) $ to obtain the empirical variables $E$
if the true model is $ M $, i.e. by the probability of Eq. \ref{probaaempi} displaying the large deviation behavior
\begin{eqnarray}
{\cal L}_T(  M \vert E ) = P_T( E \vert  M)\opsimeq_{T \to +\infty}  C(E) \  e^{\displaystyle  - T  I_{[M]} (E)    }
\label{deflikelihood}
\end{eqnarray}
The maximum of this likelihood corresponds to the vanishing of the positive rate function $ I_{[M]} (E)   $ discussed in Eq. \ref{rateempizerotyp}, i.e. the best inferred model is 
the modified model $\hat M (E)$ introduced in Eq. \ref{modeleeff}
that makes the empirical observables $E$ typical.
Via the bijective change of variables $\hat M (E) $ of Eq. \ref{modeleeff}
between the empirical data $E$ and the best inferred model $\hat M (E)$,
Eq. \ref{probaaempi}
can be translated into the large deviation form for the probability to infer the model $\hat M $ 
\begin{eqnarray}
 P^{Infer}_T ( \hat M ) && = \sum_E P_T  (E) \delta \left(\hat M - \hat M (E) \right)
  \opsimeq_{T \to +\infty}  \sum_E   C(E) \  e^{\displaystyle  - T  I_{[M]} (E)    } \delta \left(\hat M - \hat M (E) \right)
 \nonumber \\
 &&
 \opsimeq_{T \to +\infty}  {\cal C}(\hat M)  e^{\displaystyle  - T \  {\cal I}_{[M]} (\hat M )    }
\label{probainfer}
\end{eqnarray}
where the rate function $ {\cal I}_M (\hat M )$ can be explicitly obtained via the translation of 
the rate function $ I_{[M]} (E) $ at Level 2.5 of Eq. \ref{rateempi},
while the prefactor ${\cal C}(\hat M)  $ represents the translation of the constraints $C(E)$ introduced in Eq. \ref{omegat}.


\subsection{  Large deviations for inferred parameters if the true model $M$ is parametrized by a few parameters $\theta$  }

\label{sec_para}

In the previous subsection, we have described the 'full inference' problem where one considers
the best inferred model $\hat M$ that can be reconstructed from the 
full information on the relevant empirical observables $E$.
However sometimes one prefers to assume that the true model $M$ belongs to some subspace $M_{\theta}$
parametrized by a few parameters $\theta$ that one wishes to infer.
The probability to infer the parameters $\hat \theta$
is then obtained by applying Eq. \ref{probainfer} to the special case $M=M_{\theta}$ and $ \hat M = M_{\hat \theta}$
on the right handside
\begin{eqnarray}
 P^{Infer}_T ( \hat \theta ) \opsimeq_{T \to +\infty}  {\cal C}(\hat M_{\hat \theta})  e^{\displaystyle  - T \  {\cal I}_{[M_{\theta}]} ( M_{\hat \theta} )    }
\label{probainfertheta}
\end{eqnarray}


\subsection{ Application to the simplest example concerning the drawing of $T$ independent variables  }

In order to see more concretely how the general formalism described above works in practice,
it is useful to revisit now the trivial example of independent variables
before focusing on Markov models in the other sections.
In this subsection, we thus consider the problem
of drawing $T$ independent random variables $x(t)$ where $t=1,..,T$ with the discrete probability distribution $P_x$
normalized to unity
\begin{eqnarray}
  \sum_{x} P_x =1
\label{nprobaxn}
\end{eqnarray}
so that the distribution $P_.$ represents the model $M$ that one wishes to infer.

\subsubsection{ Identification of the relevant empirical observables $E$ }

The probability to draw the sequence $x(1 \leq t \leq T)$ can be rewritten in the form of Eq. \ref{ptrajectempi}
\begin{eqnarray}
 {\cal P}[x(0 \leq t \leq T)] = \prod_{t=1}^T P_{x(t) } = e^{\displaystyle  \sum_{t=1}^T \ln P_{x(t)} } 
 = e^{ \displaystyle T \sum_x \rho_x \ln P_x }
\label{pindep}
\end{eqnarray}
The only relevant empirical observable $E$ is thus the empirical density (or the empirical histogram)
\begin{eqnarray}
\rho_x \equiv  \frac{1}{T} \sum_{t=1}^T \delta_{x(t),x}
\label{empiricalhisto}
\end{eqnarray}
normalized to unity
\begin{eqnarray}
  \sum_{x} \rho_x =1
\label{nempiricalhisto}
\end{eqnarray}
The intensive action defined in Eq. \ref{ptrajectempi} is simply
\begin{eqnarray}
A_{[P]} \left( \rho_.  \right) = - \sum_x \rho_x \ln P_x 
\label{actionindep}
\end{eqnarray}

\subsubsection{ Large deviations for the empirical observable $E$, i.e. for the empirical density $\rho_.$}

The typical value of the empirical density $\rho_.$ of Eq. \ref{empiricalhisto}
is the true probability $P_.$
\begin{eqnarray}
 \rho^{typ}_{x} = P_x 
\label{rhohitotyptyp}
\end{eqnarray}
Reciprocally, the modified probability $\hat P_.$ that would make the empirical density $\rho_.$ typical
is simply
\begin{eqnarray}
\hat P_{x } = \rho_{x}
\label{indepinfer}
\end{eqnarray}
The corresponding action of Eq. \ref{actionindep} for this modified model
\begin{eqnarray}
A_{[\hat P]} \left( \rho_.  \right) =- \sum_x \rho_x \ln \hat P_x  =  - \sum_x \rho_x \ln \rho_x 
\label{actionindephat}
\end{eqnarray}
is simply the Shannon entropy of the empirical density $\rho_.$.
The probability to see the empirical density $\rho_.$ given the true probability $P_.$  
follows the Large Deviation form for large $T$
\begin{eqnarray}
P_T( \rho_. ) \opsimeq_{T \to +\infty} \delta \left(  \sum_{x} \rho_x -1 \right) e^{- T I_{[P_.]}(\rho_.) }
\label{alevel2}
\end{eqnarray}
where the normalization constraint $\left(\sum_{x} \rho_x =1 \right)$ represents the constitutive constraint denoted by $C(E)$ in Eq. \ref{probaaempi}, 
while the rate function $ I(\rho_.)$ corresponds to the difference of 
Eq. \ref{rateempi}
between the actions of Eqs \ref{actionindep} and \ref{actionindephat} 
\begin{eqnarray}
 I_{[P_.]}( \rho_. ) = A_{ [P]} ( \rho_.  ) - A_{[\hat P]} ( \rho_.  ) = \sum_{x} \rho_x \ln \frac{\rho_x}{P_x } 
  \label{rateindep}
\end{eqnarray}
This standard result is known as Sanov's Theorem :
the rate function of Eq. \ref{rateindep}
involves the relative entropy of the empirical density $\rho_.$
with respect to the true distribution $P_.$.
It is positive $I(\rho_.) \geq 0 $ and 
vanishes only for the typical value of the empirical density of Eq. \ref{rhohitotyptyp}.

\subsubsection{ Large deviations for the inferred model $\hat M$, i.e. for the inferred distribution $\hat P_.$  }

Here the best inferred distribution $\hat P$ simply coincides with the empirical distribution $\rho_.$ (Eq. \ref{indepinfer}).
As a consequence, Eq. \ref{alevel2} can be directly rephrased as the probability to infer the probability distribution 
 $\hat P_. $ from the $T$ variables drawn with the true probability $P_.$
\begin{eqnarray}
P^{Infer}_T( \hat P_. \vert P_.) \opsimeq_{T \to +\infty} \delta \left(  \sum_{x} \hat P_x -1 \right) 
e^{ \displaystyle - T \sum_{x} \hat P_x  \ln \frac{\hat P_x }{P_x }  }
\label{inferlevel2}
\end{eqnarray}

\subsubsection{ Translation for the case of continuous distribution $P(.)$ }

Up to now we have considered the case of a discrete distribution $P_.$.
However if one wishes to infer a continuous distribution $P(x)$,
one just needs to replace discrete sums by integrals in the final result of Eq. \ref{inferlevel2}
\begin{eqnarray}
P^{Infer}_T( \hat P(.) \vert P(.)) \opsimeq_{T \to +\infty} \delta \left(  \int dx \hat P (x) -1 \right) 
e^{ \displaystyle - T \int dx \hat P (x)  \ln \frac{\hat P (x) }{ P (x) }  }
\label{inferlevel2c}
\end{eqnarray}

\subsubsection{ Example of inference of two parameters only  }

As an example of the parameters inference described in subsection \ref{sec_para},
let us assume that the true distribution is the gamma distribution 
normalized on $x \in [0,+\infty[$
\begin{eqnarray}
P_{\alpha,\theta}(x) = \frac{x^{\alpha-1} }{\Gamma(\alpha) \theta^{\alpha} } e^{ - \frac{x}{\theta} }
 \label{gammadistri}
\end{eqnarray}
with the properties
\begin{eqnarray}
\int_0^{+\infty} dx x P_{\alpha,\theta}(x) && = \alpha \theta
\nonumber \\
\int_0^{+\infty} dx (\ln x) P_{\alpha,\theta}(x) && = \ln( \theta ) +  \frac{\Gamma'(\alpha)}{\Gamma(\alpha)}
 \label{gammadistriprop}
\end{eqnarray}
One considers that both 
the shape parameter $\alpha$ and the scale parameter $\theta$ are unknown and need to be inferred from data.
The probability to infer the two parameters $(\hat \alpha,\hat \theta)$
follows the large deviation of
Eq. \ref{probainfertheta} that can be evaluated using Eq. \ref{inferlevel2c}
\begin{eqnarray}
&& P^{Infer}_T( \hat \alpha,\hat \theta \vert \alpha,\theta) 
 \opsimeq_{T \to +\infty}
e^{ \displaystyle - T \int_0^{+\infty} dx P_{\hat \alpha,\hat \theta} (x)  \ln \frac{ P_{\hat \alpha,\hat \theta} (x) }{ P_{\alpha,\theta} (x) }  }
\nonumber \\
&& \opsimeq_{T \to +\infty}
e^{ - \displaystyle  T 
\left[ \ln \left( \frac{\Gamma(\alpha)\theta^{\alpha} }{\Gamma(\hat \alpha) \hat \theta^{\hat \alpha}} \right)
+ (\hat \alpha-\alpha)   \int_0^{+\infty} dx (\ln x) P_{\hat \alpha,\hat \theta} (x) 
+  \left( \frac{1}{\theta} - \frac{1}{\hat \theta}\right) \int_0^{+\infty} dx x P_{\hat \alpha,\hat \theta} (x)
\right] 
 }
\nonumber \\
&& \opsimeq_{T \to +\infty}
e^{ - \displaystyle  T 
\left[ \ln \left( \frac{\Gamma(\alpha)\theta^{\alpha} }{\Gamma(\hat \alpha) \hat \theta^{\hat \alpha}} \right)
+ (\hat \alpha-\alpha)   \left(\ln( \hat \theta ) +  \frac{\Gamma'(\hat \alpha)}{\Gamma(\hat \alpha)} \right)  
+  \left( \frac{\hat \theta}{\theta} - 1 \right) \hat \alpha 
\right] 
 }
\label{inferlevel2cgamma}
\end{eqnarray}


\section{ Inference for Markov Chain in discrete time with steady state }

\label{sec_chain}

The inference for discrete-time Markov chains has a long history in the mathematical literature \cite{1957,1961,zeitouni}.
In this section, the goal is to revisit this problem via the large deviations at Level 2.5
that have emerged more recently  \cite{fortelle_thesis,fortelle_chain,review_touchette,c_largedevdisorder,c_reset}.

\subsection{Markov chain in discrete time and discrete space parametrized by the Markov matrix $W_{.,.}$}

In this section, we consider the Markov chain dynamics for the probability $P_y(t)  $ to be at position $y$ at time $t$
\begin{eqnarray}
P_x(t+1) =  \sum_y W_{x,y}  P_y(t)
\label{markovchain}
\end{eqnarray}
where the Markov Matrix elements are positive $W_{x,y} \geq 0 $ and satisfy the normalization
\begin{eqnarray}
  \sum_x W_{x,y} && =1
\label{markovnorma}
\end{eqnarray}
So here the Markov Matrix $W_{.,.}$ represents the model $M$ that one wishes to infer.

We will assume that the steady-state solution $P^*_x \geq 0$ of Eq. \ref{markovchain}
\begin{eqnarray}
P^*_x =  \sum_y W_{x,y} P^*_y
\label{markovchainst}
\end{eqnarray}
exists. From the point of view of the Perron–Frobenius theorem,
Eqs \ref{markovnorma} and \ref{markovchainst} mean that unity is the highest eigenvalue of the positive Markov Matrix $W_{.,.}$,
where the positive left eigenvector $l_x$ is constant
\begin{eqnarray}
 l_x=1
\label{markovleft}
\end{eqnarray}
while  the right eigenvector $r_x$ is the steady state
\begin{eqnarray}
 r_x=P^*_x
\label{markovright}
\end{eqnarray}


\subsection{ Identification of the relevant time-empirical observables that determine the trajectories probabilities }

The probability of the whole trajectory $x(0 \leq t \leq T)$ starting at the fixed position $x_0$ at time $t=0$
\begin{eqnarray}
{\cal P}[x(0 \leq t \leq T)]  =   \delta_{x(0),0 } \left[ \prod_{t=1}^T W_{x(t) ,x(t-1)} \right] 
=   \delta_{x(0),0 } e^{ \displaystyle \sum_{t=1}^T \ln \left(W_{x(t) ,x(t-1)} \right) }
\label{pwtraj}
\end{eqnarray}
can be rewritten in terms of the time-empirical 2-point density that characterizes the flows between two consecutive positions
within this trajectory  $x(0 \leq t \leq T)$
\begin{eqnarray}
 \rho^{(2)}_{x,y} && \equiv \frac{1}{T} \sum_{t=1}^T \delta_{x(t),x}  \delta_{x(t-1),y} 
\label{rho2pt}
\end{eqnarray}
as
\begin{eqnarray}
{\cal P}[x(0 \leq t \leq T)]  =  \delta_{x(0),0 } \  e^{ \displaystyle T \sum_{x,y}  \rho^{(2)}_{x,y} \ln \left(W_{x ,y} \right) } 
\label{pwtrajempi}
\end{eqnarray}

With respect to the general formalism summarized in Section \ref{sec_general},
this means that the relevant empirical observable $E$ is the 2-point density $\rho^{(2)}_{.,.}$ of Eq. \ref{rho2pt},
and that the intensive action introduced in Eq. \ref{ptrajectempi} reads
\begin{eqnarray}
A_{[W]} \left( \rho^{(2)}_{.,.} \right) =- \sum_{x,y}  \rho^{(2)}_{x,y} \ln \left(W_{x ,y} \right)
\label{actionchain}
\end{eqnarray}
Note that the 2-point density of Eq. \ref{rho2pt} contains the information on the empirical 1-point density
that can be obtained via the sum over the first or the second position 
 (up to a boundary term of order $1/T$ that is negligible for large duration $T \to +\infty$)
\begin{eqnarray}
 \rho_{x} && \equiv \frac{1}{T} \sum_{t=1}^T \delta_{x(t),x}  = \sum_y \rho^{(2)}_{x,y} =
 \sum_y \rho^{(2)}_{y,x}
\label{rho1pt}
\end{eqnarray}
with the normalization
\begin{eqnarray}
\sum_x \rho_{x} && = 1
\label{rho1ptnorma}
\end{eqnarray}


\subsection{ Typical values of the empirical observables } 

The typical value of the empirical 1-point density is the steady state of Eq. \ref{markovchainst}
\begin{eqnarray}
 \rho^{typ}_{x} = P^*_x 
\label{rho1pttyp}
\end{eqnarray}
while the typical value of the empirical 2-point density is given by the steady-state flows
\begin{eqnarray}
 \rho^{(2)typ}_{x,y} = W_{x,y}  P^*_y  
\label{rho2pttyp}
\end{eqnarray}

Reciprocally, the modified Markov matrix elements $\hat W_{.,.}$ that would make the empirical densities 
$\rho_.$ and $\rho^{(2)}_{.,.}$ typical
are given by the following ratios 
\begin{eqnarray}
\hat W_{x ,y} \equiv \frac{\rho^{(2)}_{x,y}}{ \rho_{y}} 
\label{Winfer}
\end{eqnarray}

With respect to the general formalism summarized in Section \ref{sec_general},
this means that the intensive action of Eq. \ref{actionchain} reads for the modified Markov matrix $\hat W_{.,.}$ of Eq. \ref{Winfer}
\begin{eqnarray}
A_{[\hat W]} \left( \rho^{(2)}_{.,.} \right) =- \sum_{x,y}  \rho^{(2)}_{x,y} \ln \left( \frac{\rho^{(2)}_{x,y}}{ \rho_{y}}   \right)
\label{actionchaintyp}
\end{eqnarray}


\subsection{ Large deviations at level 2.5 for the relevant time-empirical observables  }

For large $T$, the joint probability to see the empirical 2-point and 1-point densities 
follows the large deviation form at Level 2.5  \cite{fortelle_thesis,fortelle_chain,review_touchette,c_largedevdisorder,c_reset}
\begin{eqnarray}
P_T ( \rho^{(2)}_{.,.} ; \rho_. )
 && \opsimeq_{T \to +\infty}  C ( \rho^{(2)}_{.,.} ; \rho_.)e^{ - T I_{2.5}( \rho^{(2)}_{.,.} ; \rho_.)  } 
\label{proba2.5chain}
\end{eqnarray}
with the constraints discussed in Eqs \ref{rho1pt} and \ref{rho1ptnorma}
\begin{eqnarray}
&& C ( \rho^{(2)}_{.,.} ; \rho_.)
  = \delta \left( \sum_x \rho_{x} - 1 \right) 
  \left[ \prod_x \delta \left(   \sum_y \rho^{(2)}_{x,y} -\rho_{x} \right) \right]
 \left[ \prod_y \delta \left(   \sum_x \rho^{(2)}_{x,y} -\rho_{y} \right) \right]
\label{constraints2.5chain}
\end{eqnarray}
while the rate function 
\begin{eqnarray}
  I_{2.5}( \rho^{(2)}_{.,.} ; \rho_. ) 
  =  
\sum_x \sum_y \rho^{(2)}_{x,y} \ln \left( \frac{\rho^{(2)}_{x,y}}{ W_{x,y}  \rho_{y}}  \right) 
\label{rate2.5chain}
\end{eqnarray}
is positive and vanishes only for the typical values of Eqs \ref{rho1pttyp} and \ref{rho2pttyp}.
With respect to the general formalism summarized in Section \ref{sec_general},
the rate function of Eq. \ref{rate2.5chain}
indeed corresponds to the difference of Eq. \ref{rateempi}
between the actions of Eqs \ref{actionchain} and \ref{actionchaintyp} 
\begin{eqnarray}
 I_{2.5}( \rho^{(2)}_{.,.} ; \rho_. ) = A_{[ W]} \left( \rho^{(2)}_{.,.} \right)  - A_{[\hat W]} \left( \rho^{(2)}_{.,.} \right) 
 \label{ratefromactionchain}
\end{eqnarray}


\subsection{ Probability to infer the Markov matrix $\hat W_{.,.}$ with its associated steady state $\hat P^*_.$}

From the empirical 1-point and 2-point densities measured over a very long trajectory $x(0 \leq t \leq T)$
(see Eqs \ref{rho2pt} and \ref{rho1pt}),
the maximum likelihood inference yields that 
the best inferred steady state $P^*_.$ corresponds to the 1-point empirical density (see Eq. \ref{rho1pttyp})
\begin{eqnarray}
\hat P_x^* = \rho_x
\label{pstinferchain}
\end{eqnarray}
while the best inferred Markov matrix $\hat W_{.,.}$ corresponds to the modified matrix of Eq. \ref{Winfer}.

Via this bijective change of variables, Eq. \ref{proba2.5chain} yields that the joint probability to infer 
the Markov matrix $\hat W_{.,.}$ and its associated steady state $\hat P^*_.$ reads
\begin{eqnarray}
P_T^{Infer} ( \hat W_{.,.} ; \hat P^*_. )
 &&
  \opsimeq_{T \to +\infty}  
  \delta \left( \sum_x \hat P^*_{x} - 1 \right) 
 \left[  \prod_x \delta \left(   \sum_y \hat W_{x ,y} \hat P^*_{y}-\hat P^*_{x} \right) \right]
 \left[ \prod_y \delta \left(   \sum_x \hat W_{x ,y} -1 \right) \right]
 \nonumber \\
&&  e^{ \displaystyle - T \sum_y \hat P^*_{y} \sum_x  \hat W_{x ,y}  \ln \left( \frac{\hat W_{x ,y} }{ W_{x,y}  }  \right) 
 } 
\label{proba2.5chaininfer}
\end{eqnarray}
The first constraint corresponds to the normalization of the inferred steady state $\hat P^* $.
The two other constraints mean that the inferred Markov Matrix $\hat W_{.,.}$
has unity as highest eigenvalue, with the inferred steady state $\hat r_. = \hat P^*_. $ as right eigenvector,
and the trivial left eigenvector $\hat l_x = 1$. So these constraints are in direct correspondence with 
the properties of the true steady state and the true Markov Matrix (see Eqs \ref{markovnorma}
and \ref{markovchainst}).

The formula of Eq. \ref{proba2.5chaininfer} will be applied in section \ref{sec_RWRing}
to the random walk on a disordered ring.


\subsection{ Translation for discrete-time Markov chains in continuous space $\vec x$ in dimension $d$ }

For discrete-time Markov chains in continuous space $\vec x$ in dimension $d$ with kernel $W(\vec x, \vec y)  $
\begin{eqnarray}
P_{t+1}( \vec x) =  \int d^d \vec y  \ W(\vec x, \vec y)  P_{t}( \vec y) 
\label{markovchainc}
\end{eqnarray}
one just needs to replace discrete sums by integrals in the final result of Eq. \ref{proba2.5chaininfer}
to obtain that
the joint probability to infer 
the Markov kernel $\hat W(.,.)$ with its associated steady state $\hat P^*(.)$
is given by
\begin{eqnarray}
&& P_T^{Infer} ( \hat W(.,.) ; \hat P^*(.) )
  \opsimeq_{T \to +\infty}  
  \delta \left( \int d^d \vec x \  \hat P^*( \vec x ) - 1 \right) 
  \left[ \prod_{\vec x} \delta \left(    \int d^d \vec y \  \hat W(\vec x, \vec y)  \hat P^*( \vec y) - \hat P^*( \vec x)\right) \right]
 \left[ \prod_y \delta \left(  \int d^d \vec x  \ \hat W(\vec x, \vec y)  -1 \right) \right]
 \nonumber \\
&&  e^{ \displaystyle - T  \int d^d \vec y  \ \hat P^*( \vec y)
  \int d^d \vec x \   \hat W(\vec x, \vec y)  \ln \left( \frac{ \hat W(\vec x, \vec y) }{  W(\vec x, \vec y)  }  \right) 
 } 
\label{proba2.5chaininfercontinuous}
\end{eqnarray}
This formula will be useful in subsection \ref{subsec_ito}.


\section{  Application to the Random Walk on the disordered ring of $L$ sites }

\label{sec_RWRing}

In this section, the large deviations analysis of inference for discrete-time Markov chains described in the previous section
is applied to the example of the random walk on a disordered ring
\cite{Der_Pom,derrida}.

\subsection{ Model parametrization and non-equilibrium steady state }

The Derrida-Pomeau model \cite{Der_Pom,derrida}
is defined on a ring of $L$ sites with periodic boundary conditions $x+L \equiv x$,
and corresponds to the dynamics of Eq. \ref{markovchain}
where the
Markov Matrix
\begin{eqnarray}
 W_{x,y}  = \delta_{x,y+1} R_y + \delta_{x,y-1} (1-R_y)
\label{ringpomeau}
\end{eqnarray}
is parametrized by the $L$ parameters $R_y\in ]0,1[$ for $y=1,..,L$.
So when the particle is on site $y$ at time $t$,
the new position at time $(t+1)$ can be either the right neighbor $(y+1)$ with probability $R_y \in ]0,1[$
or the left neighbor $(y-1)$ with the complementary probability $(1-R_y) \in ]0,1[$.

The steady state of Eq. \ref{markovchainst}
\begin{eqnarray}
P^*_x =  R_{x-1} P^*_{x-1} +  (1-R_{x+1}) P^*_{x+1}
\label{markovchainstring}
\end{eqnarray}
 reads \cite{Der_Pom,derrida}
\begin{eqnarray}
 P_x^* 
 = \frac{K} {R_x} \left[ 1+ \sum_{z=1}^{L-1} \prod_{y=1}^z s_{x+y} \right]
  = \frac{K} {R_x} \left[ 1+s_{x+1}+s_{x+1}s_{x+2} + ... + s_{x+1}s_{x+2}... s_{x+L-1} \right]
\label{ringst}
\end{eqnarray}
in terms of the ratios
\begin{eqnarray}
s_x  \equiv \frac{1-R_x} {R_x} 
\label{ratiokesten}
\end{eqnarray}
while the constant $K$ is fixed by the normalization
\begin{eqnarray}
1=\sum_{x=1}^L P_x^* 
 = K  \sum_{x=1}^L \frac{1} {R_x} \left[ 1+ \sum_{z=1}^{L-1} \prod_{y=1}^z s_{x+y} \right]
\label{ringstnorma}
\end{eqnarray}

When the probabilities $R_y$ are random, the characteristic structure of Eq. \ref{ringst} is 
known as Kesten random variables 
and appears in many disordered systems
\cite{Kesten,Solomon,sinai,jpb_review,Der_Hil,Cal,strong_review,c_microcano,c_watermelon,c_mblcayley,c_reset}.
The generalization to the matrix framework is discussed 
in the recent work \cite{pldjbp} and in references therein.


\subsection{ Inference of the $L$ parameters $R_y$ of the model  }

Here the trajectory data are the positions $x(t)$ for the discrete times $t=0,1,2,..,T$.
For each site $y=1,..,L$ on the ring, one computes the 2-point density $\rho^{(2)}_{x,y}$ of Eq. \ref{rho2pt}
for the only two possible values $x=y \pm 1$ for the model of Eq. \ref{ringpomeau}
\begin{eqnarray}
 \rho^{(2)}_{y +1,y} && = \frac{1}{T} \sum_{t=1}^T \delta_{x(t),y+1}  \delta_{x(t-1),y} 
 \nonumber \\
 \rho^{(2)}_{y -1,y} && = \frac{1}{T} \sum_{t=1}^T \delta_{x(t),y-1}  \delta_{x(t-1),y} 
\label{rho2ptring}
\end{eqnarray}
The 1-point density $\rho_y$ of Eq \ref{rho1pt} corresponds to their sum
\begin{eqnarray}
 \rho_{y} = \frac{1}{T} \sum_{t=1}^T \delta_{x(t-1),y}  =  \rho^{(2)}_{y +1,y}  +  \rho^{(2)}_{y -1,y} 
 \label{rho1ptring}
\end{eqnarray}
So the non-vanishing matrix elements of the best inferred Markov matrix $\hat W_{x,y}$ of Eq. \ref{Winfer} for $x=y \pm 1$
are computed via the ratios
\begin{eqnarray}
\hat W_{y+1 ,y} && \equiv \frac{\rho^{(2)}_{y+1,y}}{ \rho_{y}} = \frac{\rho^{(2)}_{y+1,y}}{ \rho^{(2)}_{y +1,y}  +  \rho^{(2)}_{y -1,y} }
\equiv \hat R_y
\nonumber \\
\hat W_{y-1 ,y} && \equiv \frac{\rho^{(2)}_{y-1,y}}{ \rho_{y}} =  \frac{\rho^{(2)}_{y-1,y}}{\rho^{(2)}_{y +1,y}  +  \rho^{(2)}_{y -1,y} } 
\equiv 1- \hat R_y
\label{Winferring}
\end{eqnarray}
In summary, the trajectory data have been used to compute the $L$ parameters $\hat R_y\in ]0,1[$
that parametrize the best inferred Markov matrix $\hat W_{.,.}$ of Eq. \ref{Winfer} 
\begin{eqnarray}
 \hat W_{x,y}  = \delta_{x,y+1} \hat R_y + \delta_{x,y-1} (1- \hat R_y)
\label{ringpomeauinfer}
\end{eqnarray}


\subsection{ Large deviations for the $L$ inferred parameters $\hat R_y$  }

Now one wishes to know how the $L$ inferred parameters $\hat R_y\in ]0,1[$ computed from the data
can fluctuate with respect to the 'true' values $R_y$ of the 'true' model of Eq. \ref{ringpomeau}.
The inferred steady state $\hat P_x^*$ of Eq. \ref{pstinferchain} corresponds to the steady state associated to the
model with the inferred parameters $ \hat R_y $ and is thus given by the analog of Eqs \ref{ringst} and \ref{ratiokesten}
\begin{eqnarray}
 \hat P_x^* 
 = \frac{ \hat K} { \hat R_x} \left[ 1+ \sum_{z=1}^{L-1} \prod_{y=1}^z \left( \frac{1- \hat R_{x+y}} { \hat R_{x+y}}  \right)  \right]
\label{ringstinfer}
\end{eqnarray}
where the constant $\hat K$ is fixed by the normalization as in Eq. \ref{ringstnorma}
\begin{eqnarray}
1=\sum_{x=1}^L \hat P_x^* 
 = \hat K  \sum_{x'=1}^L  \frac{ 1 } { \hat R_{x'}} \left[ 1+ \sum_{z=1}^{L-1} \prod_{y=1}^z \left( \frac{1- \hat R_{x'+y}} { \hat R_{x'+y}}  \right)  \right]
 \label{ringstnormainfer}
\end{eqnarray}

As a consequence, 
the joint probability to infer 
the $L$ parameters $\hat R_.$ of a given disordered ring
follows the large deviation form of Eq. \ref{proba2.5chaininfer} 
\begin{eqnarray}
P_T^{Infer} ( \hat R_. )
  \opsimeq_{T \to +\infty}  
   e^{  - T {\cal I}_{R_.} ( \hat R_. ) } 
\label{proba2.5chaininferRing}
\end{eqnarray}
with the explicit rate function
\begin{eqnarray}
{\cal I}_{R_.} ( \hat R_. )
&& = \sum_{x=1}^L 
\left[   \hat R_x \ln \left( \frac{   \hat R_x }{ R_x  }  \right) 
+ (1- \hat R_x)  \ln \left( \frac{ 1- \hat R_x }{ 1-R_x  }  \right) 
\right]
\hat P^*_x
\nonumber \\
&& 
=\sum_{x=1}^L 
\left[   \hat R_x \ln \left( \frac{   \hat R_x }{ R_x  }  \right) 
+ (1- \hat R_x)  \ln \left( \frac{ 1- \hat R_x }{ 1-R_x  }  \right) 
\right]
\frac{\displaystyle \frac{1} { \hat R_x} \left[ 1+ \sum_{z=1}^{L-1} \prod_{y=1}^z \left( \frac{1- \hat R_{x+y}} { \hat R_{x+y}}  \right)  \right]}
{\displaystyle\sum_{x'=1}^L  \frac{ 1 } { \hat R_{x'}} \left[ 1+ \sum_{z=1}^{L-1} \prod_{y=1}^z \left( \frac{1- \hat R_{x'+y}} { \hat R_{x'+y}}  \right)  \right]}
\label{rateinferRing}
\end{eqnarray}
The main qualitative conclusion is thus that the $L$ inferred parameters $  \hat R_x$
are coupled via the inferred steady state $\hat P^*_. $ that they produce together.


\section{ Inference for Markov jump process in continuous time with steady state  }

\label{sec_jump}

The inference for continuous-time jump processes has been analyzed in various contexts
including evolution models
\cite{evolution}, DNA unzipping \cite{monassonDNAprl,monassonDNApre}
and the continuous-time random walk in a one-dimensional disordered medium \cite{monassonsinai2008,monassonsinai2009}.
In this section, the goal is to revisit this problem via the large deviations at Level 2.5
that have emerged more recently \cite{fortelle_thesis,fortelle_jump,maes_canonical,maes_onandbeyond,wynants_thesis,chetrite_formal,BFG1,BFG2,chetrite_HDR,c_ring,c_interactions,c_open,barato_periodic,chetrite_periodic,c_reset}.

\subsection{  Markov jump process in continuous time and discrete space  }

In this section, we consider the continuous-time dynamics in discrete space defined by the Master Equation
\begin{eqnarray}
\frac{\partial P_x(t)}{\partial t} =    \sum_{y }   w_{x,y}  P_y(t) 
\label{mastereq}
\end{eqnarray}
where the off-diagonal $x \ne y$ positive matrix elements $w_{x,y} \geq 0 $  represent the transitions rates 
per unit time from $y$ to $x $,
while the corresponding diagonal elements are negative and fixed by the conservation of probability to be
\begin{eqnarray}
w_{y,y}  && =  - \sum_{x \ne y} w_{x,y} 
\label{wdiag}
\end{eqnarray}
As in Eq. \ref{markovchainst}, we will assume that the steady-state $P^*_x$ of Eq. \ref{mastereq}
\begin{eqnarray}
0 =    \sum_{y }   w_{x,y}  P_y^* 
\label{mastereqst}
\end{eqnarray}
exists.
Eqs \ref{wdiag} and \ref{mastereqst} mean that zero is the highest eigenvalue of the Markov Matrix $w_{.,.}$,
with the positive left eigenvector 
\begin{eqnarray}
 l_x=1
\label{markovleftj}
\end{eqnarray}
and the positive right eigenvector $r_x$ given by the steady state
\begin{eqnarray}
 r_x=P^*_x
\label{markovrightj}
\end{eqnarray}


\subsection{Identification of the relevant time-empirical observables that determine the trajectories probabilities }

The probability of the whole trajectory $x(0 \leq t \leq T)$ 
\begin{eqnarray}
{\cal P}[x(0 \leq t \leq T)]   
=  e^{ \displaystyle  \left[  \sum_{t: x(t^+) \ne x(t^-) } \ln ( w_{x(t^+) , x(t^-) } ) +  \int_0^T dt  w_{x(t) , x(t) }   \right]  }
\label{pwtrajjump}
\end{eqnarray}
can be rewritten as
\begin{eqnarray}
{\cal P}[x(0 \leq t \leq T)]   
=  e^{ \displaystyle  T \left[  \sum_{y  }\sum_{ x \ne y  }
q_{x,y}\ln ( w_{x,y } ) + \sum_x     \rho_{x}w_{x , x }     \right]
}
\label{pwtrajjumpempi}
\end{eqnarray}
in terms of the empirical density
\begin{eqnarray}
 \rho_{x} && \equiv \frac{1}{T} \int_0^T dt \  \delta_{x(t),x}  
 \label{rho1pj}
\end{eqnarray}
normalized to unity
\begin{eqnarray}
\sum_x \rho_{x} && = 1
\label{rho1ptnormaj}
\end{eqnarray}
and in terms of the jump densities for $x \ne y$
\begin{eqnarray}
q_{x,y} \equiv  \frac{1}{T} \sum_{t : x(t^+) \ne x(t^-)} \delta_{x(t^+),x} \delta_{x(t^-),y} 
\label{jumpempiricaldensity}
\end{eqnarray}
that should satisfy the stationarity constraint (for any $x$, the total incoming flow should be equal to the total outgoing flow)
\begin{eqnarray}
\sum_{y \ne x} q_{x,y} = \sum_{y \ne x} q_{y,x}
\label{contrainteq}
\end{eqnarray}

With respect to the general formalism summarized in Section \ref{sec_general},
this means that the relevant empirical observables $E$ are the empirical density $\rho_.$ and the flows $q_{.,.}$.
The corresponding intensive action introduced in Eq. \ref{ptrajectempi} reads using Eq. \ref{wdiag}
\begin{eqnarray}
A_{[w]} \left( \rho_. ; q_{.,.} \right) && = - \sum_y     \rho_{y}w_{y , y }    -  \sum_{y  }\sum_{ x \ne y  } q_{x,y}\ln ( w_{x,y } ) 
 =   \sum_y     \sum_{x \ne y} \left[ w_{x,y}   \rho_{y}    -   q_{x,y}\ln ( w_{x,y } ) \right]
\label{actionjump}
\end{eqnarray}


\subsection{ Typical values of the empirical observables }

The typical value of the empirical density is the steady state of Eq. \ref{mastereqst}
\begin{eqnarray}
 \rho^{typ}_{x} = P^*_x 
\label{rhotypmaster}
\end{eqnarray}
while the typical value of the jump densities read
\begin{eqnarray}
 q^{typ}_{x,y} = w_{x,y}  P^*_y  
\label{qtyp}
\end{eqnarray}
Reciprocally, the modified off-diagonal $x \ne y$ matrix elements $\hat w_{x,y}$ 
that would make typical the empirical density $\rho_y$ and the empirical jump-density $q_{x,y}$
correspond to the following ratios
\begin{eqnarray}
\hat w_{x ,y} \equiv \frac{q_{x,y}}{ \rho_{y}} 
\label{winfer}
\end{eqnarray}
The corresponding modified diagonal matrix elements  
are fixed by the conservation of probabilities as in Eq. \ref{wdiag}
\begin{eqnarray}
\hat w_{y,y}  && =  - \sum_{x \ne y} \hat w_{x,y} = - \sum_{x \ne y}\frac{q_{x,y}}{ \rho_{y}} 
\label{wdiaghat}
\end{eqnarray}
With respect to the general formalism summarized in section \ref{sec_general},
this means that the intensive action of Eq. \ref{actionjump} reads for the modified Markov matrix $\hat w_{.,.}$ 
of Eqs \ref{winfer} and \ref{wdiaghat}
\begin{eqnarray}
A_{[\hat w]} \left( \rho_. ; q_{.,.} \right) = 
  \sum_y     \sum_{x \ne y} \left[ q_{x,y}    -   q_{x,y}\ln \left( \frac{q_{x,y}}{ \rho_{y}}  \right) \right]
\label{actionjumptyp}
\end{eqnarray}


\subsection{ Large deviations at level 2.5 for the empirical density and the empirical flows  }

The joint probability distribution of the empirical density $\rho_.$ and flows $q_{.,.}$
satisfy the following
large deviation form at level 2.5 
\cite{fortelle_thesis,fortelle_jump,maes_canonical,maes_onandbeyond,wynants_thesis,chetrite_formal,BFG1,BFG2,chetrite_HDR,c_ring,c_interactions,c_open,barato_periodic,chetrite_periodic,c_reset,c_runandtumble}
\begin{eqnarray}
P_{T}[ \rho_. ; q_{.,.} ] \oppropto_{T \to +\infty} C[ \rho_. ; q_{.,.} ] e^{- T I_{2.5}[ \rho_. ; q_{.,.} ] }
\label{level2.5master}
\end{eqnarray}
with the constraints discussed in Eqs \ref{rho1ptnormaj} and \ref{contrainteq}
\begin{eqnarray}
&& C [ \rho_. ; q_{.,.} ]
  = \delta \left( \sum_x \rho_{x} - 1 \right) 
 \prod_x \delta \left( \sum_{y \ne x} q_{x,y} - \sum_{y \ne x} q_{y,x}  \right)
\label{constraints2.5master}
\end{eqnarray}
while the rate function reads
\begin{eqnarray}
I_{2.5}[ \rho_. ; q_{.,.} ]=  \sum_{y } \sum_{x \ne y} 
\left[ q_{x,y}  \ln \left( \frac{ q_{x,y}  }{  w_{x,y}  \rho_y }  \right) 
 - q_{x,y}  + w_{x,y}  \rho_y  \right]
\label{rate2.5master}
\end{eqnarray}
and vanishes only for the typical values of Eqs \ref{rhotypmaster} and \ref{qtyp}.
With respect to the general formalism summarized in section \ref{sec_general},
the rate function of Eq. \ref{rate2.5master}
indeed corresponds to the difference of Eq. \ref{rateempi}
between the actions of Eqs \ref{actionjump} and \ref{actionjumptyp} 
\begin{eqnarray}
 I_{2.5}( \rho_. ; q_{.,.}  ) = A_{[ w]} \left(  \rho_. ; q_{.,.}  \right)  - A_{[\hat w]} \left(  \rho_. ; q_{.,.}  \right) 
 \label{ratefromactionjump}
\end{eqnarray}


\subsection{ Probability to infer the Markov matrix $\hat w_{.,.}$ with its associated steady state $\hat P^*_.$}

From the empirical density and the empirical flows measured over a very long trajectory $x(0 \leq t \leq T)$
(see Eqs \ref{rho1pj} and \ref{jumpempiricaldensity}),
the maximum likelihood inference yields that 
the best inferred steady state $P^*_.$ corresponds to the empirical density (see Eq. \ref{rhotypmaster})
\begin{eqnarray}
\hat P_x^* = \rho_x
\label{pstinferjump}
\end{eqnarray}
while the best inferred Markov matrix $\hat w_{.,.}$ is given by Eqs \ref{winfer} and \ref{wdiaghat}.
Via this bijective change of variables, Eq. \ref{level2.5master} yields that the joint probability to infer 
the Markov matrix $\hat w_{.,.}$ and its associated steady state $\hat P^*_.$
follows the large deviation form
\begin{eqnarray}
P_{T}^{Infer}[ \hat w_{.,.} ; \hat P^*_. ] \oppropto_{T \to +\infty} 
&& \delta \left( \sum_x \hat P^*_{x} - 1 \right) 
  \left[ \prod_x\delta \left(   \sum_y \hat w_{x ,y} \hat P^*_{y} \right) \right]
 \left[ \prod_y \delta \left(   \sum_x \hat w_{x ,y}  \right)  \right]
\nonumber \\
&&  e^{ \displaystyle - T  \sum_{y } \hat P^*_y
\sum_{x \ne y} 
\left[  \hat w_{x ,y}   \ln \left( \frac{  \hat w_{x ,y}   }{  w_{x,y}   }  \right) 
 -  \hat w_{x ,y}  + w_{x,y}    \right] }
\label{level2.5masterinfer}
\end{eqnarray}
The first constraint corresponds to the normalization of the inferred steady state $\hat P^* $.
The two other constraints mean that the inferred Markov Matrix 
has zero as highest eigenvalue, with the inferred steady state $\hat r_. = \hat P^*_. $ as right eigenvector,
and the trivial left eigenvector $\hat l_x = 1$.
Again these constraints are in direct correspondance with 
the properties of the true steady state and the true Markov Matrix (see Eqs \ref{wdiag}
and \ref{mastereqst}).

The formula of Eq. \ref{level2.5masterinfer} will be applied in section \ref{sec_TrapRing}
to the directed trap model on a disordered ring.

\subsection{ Translation for continuous-time Markov jump processes in continuous space $\vec x$ in dimension $d$ }

For continuous-time Markov jump processes in continuous space $\vec x$ in dimension $d$ with kernel $w(\vec x, \vec y)  $
\begin{eqnarray}
\frac{\partial P_t(\vec x) }{\partial t} =    \int d^d y  \  w(\vec x, \vec y)  P_t(\vec y) -  \left( \int d^d y \   w(\vec y, \vec x) \right) P_t(\vec x)
\label{mastereqc}
\end{eqnarray}
Eq. \ref{level2.5masterinfer}
translates into the following probability to infer 
the Markov kernel $\hat w(.,.)$ with its associated steady state $\hat P^*(.)$
\begin{eqnarray}
P_{T}[ \hat w(.,.) ; \hat P^*(.) ] \oppropto_{T \to +\infty} 
&& \delta \left( \int d^d y \hat P^*(\vec x) - 1 \right) 
 \prod_x  
\delta \left(   \int d^d y  \  w(\vec x, \vec y)  P^*(\vec y) -  \left( \int d^d y \   w(\vec y, \vec x) \right) P^*(\vec x)
  \right)
\nonumber \\
&&  e^{ \displaystyle - T  \int d^d y \hat P^*(\vec y) 
\int d^d x 
\left[  \hat w(\vec x, \vec y)   \ln \left( \frac{  \hat w(\vec x, \vec y)   }{  w(\vec x, \vec y)   }  \right) 
 -  \hat w(\vec x, \vec y)  + w(\vec x, \vec y)   \right] }
\label{level2.5jumpinfer}
\end{eqnarray}


\section{ Application to the Directed Trap Model on a disordered ring of $L$ sites }

\label{sec_TrapRing}

In this section, the large deviations analysis of inference for 
continuous-time Markov jump processes
 described in the previous section
is applied to the directed disordered trap model 
\cite{jpdir,aslangul,comptejpb,directedtrapandsinai},
whose large deviations properties have been studied recently \cite{vanwijland_trap,c_ring,c_ruelle}.
Note that besides this directed disordered trap model, many other trap models
have been also under study in the context of anomalously slow glassy behaviors 
\cite{jpb_weak,dean,jp_pheno,bertinjp1,bertinjp2,trapsymmetric,trapnonlinear,trapreponse,trap_traj}.

\subsection{ Model parametrization and steady state }

In this section, we consider the directed disordered trap model 
\cite{jpdir,aslangul,comptejpb,directedtrapandsinai,vanwijland_trap,c_ring,c_ruelle}
on a ring of $L$ sites with periodic boundary conditions $x+L \equiv x$.
The dynamics is defined by the master equation \ref{mastereq}
where
the Markov matrix
\begin{eqnarray}
 w_{x,y}  = \delta_{x,y+1} \frac{1}{\tau_y}  - \delta_{x,y} \frac{1}{\tau_y}
\label{directedtrap}
\end{eqnarray}
is parametrized by the $L$ trapping times $\tau_y \in ]0,+\infty[$.
So when the particle is on site $y$ at time $t$,
the only possible move is to jump to the right neighbor $(y+1)$ with the rate $\frac{1}{\tau_y} $ per unit time.

The solution for the steady state of Eq. \ref{mastereqst}
\begin{eqnarray}
0 =    \frac{1}{\tau_{x-1}}    P_{x-1}^* -   \frac{1}{\tau_x}    P_x^*
\label{trapst}
\end{eqnarray}
is simply given by
\begin{eqnarray}
 P_x^*  = \frac{ \tau_x } { \displaystyle \sum_{x'=1}^{L} \tau_{x'}}
\label{trapstsol}
\end{eqnarray}


\subsection{ Inference of the $L$ trapping times $\tau_y$ of the model  }

Here the trajectory data are the positions $x(t)$ for the continuous time $t \in [0,T]$.
For each site $y=1,..,L$ on the ring, one computes the empirical density $\rho_y$ of Eq. \ref{rho1pj},
and the empirical density of jumps $q_{x,y}$ of Eq \ref{jumpempiricaldensity} towards the only possible site $x=y+1$
in the present directed trap model of Eq. \ref{directedtrap}
\begin{eqnarray}
q_{y+1,y} =  \frac{1}{T} \sum_{t : x(t^+) \ne x(t^-)} \delta_{x(t^+),y+1} \delta_{x(t^-),y} 
\label{jumpempiricaldensitytrap}
\end{eqnarray}
So the non-vanishing off-diagonal matrix elements of the best inferred Markov matrix $\hat w_{x,y}$ for $x=y + 1$
are computed via the ratios of Eq. \ref{winfer}
\begin{eqnarray}
\hat w_{y+1 ,y} = \frac{q_{y+1,y}}{ \rho_{y}} = \frac{ \displaystyle
\sum_{t : x(t^+) \ne x(t^-)} \delta_{x(t^+),y+1} \delta_{x(t^-),y} }{\int_0^T dt \  \delta_{x(t),y}} \equiv \frac{1}{\hat \tau_y} 
\label{winfertrap}
\end{eqnarray}
and provide the $L$ inferred trapping times $ \hat \tau_y$.
The corresponding diagonal matrix elements are given by Eq. \ref{wdiaghat}
\begin{eqnarray}
\hat w_{y,y}   =  -  \hat w_{y+1,y} = -  \frac{1}{\hat \tau_y} 
\label{wdiaghattrap}
\end{eqnarray}
In summary, the trajectory data have been used to compute the $L$ parameters $ \hat \tau_y$
that parametrize the best inferred Markov matrix  
\begin{eqnarray}
 \hat w_{x,y}  = \delta_{x,y+1} \frac{1}{\hat \tau_y}  - \delta_{x,y} \frac{1}{\hat \tau_y}
\label{trapinfer}
\end{eqnarray}


\subsection{ Large deviations for the $L$ inferred parameters $\hat \tau_y$  }

Now one wishes to know how the $L$ inferred parameters $\hat \tau_y$ computed from the data
can fluctuate with respect to the 'true' values $\tau_y$ of the 'true' model of Eq.  \ref{directedtrap}.
The inferred steady state $\hat P_x^*$ corresponds to the steady state associated to the
model with the inferred parameters $ \hat \tau_y $ and is thus given by the analog of Eq. \ref{trapst} 
\begin{eqnarray}
 \hat P_x^*  =  \frac{ \hat \tau_x } { \displaystyle \sum_{x'=1}^{L} \hat  \tau_{x'}}
\label{trapsttinfer}
\end{eqnarray}
As a consequence, 
the joint probability to infer 
the $L$ parameters $\hat \tau_.$ of a given disordered ring
follows the large deviation form
of Eq. \ref{level2.5masterinfer} 
\begin{eqnarray}
P^{Infer}_T ( \hat \tau_. )
  \opsimeq_{T \to +\infty}  
   e^{  - T {\cal I}_{\tau_.} ( \hat \tau_. ) } 
\label{proba2.5ntrapinfer}
\end{eqnarray}
with the explicit rate function
\begin{eqnarray}
{\cal I}_{\tau_.} ( \hat \tau_. )  =  \sum_{x=1}^L  \hat P^*_x
\left[   \frac{1}{\hat \tau_x}   \ln \left( \frac{   \tau_x   }{  \hat \tau_x }  \right) 
 -   \frac{1}{\hat \tau_x}  + \frac{1}{ \tau_x}    \right] 
 =   \frac{ \displaystyle \sum_{x=1}^L 
\left[ -   \ln \left( \frac{  \hat \tau_x }{   \tau_x   }  \right) 
 -   1  + \frac{ \hat \tau_x}{ \tau_x}    \right]  } 
 { \displaystyle \sum_{x'=1}^{L} \hat  \tau_{x'}}
\label{level2.5traprinfer}
\end{eqnarray}
As in Eq. \ref{rateinferRing},
the main qualitative conclusion is that the $L$ inferred parameters $  \hat \tau_x$
are coupled via the inferred steady state $\hat P^*_. $ that they produce together.


\section{ Inference for Diffusion Processes in dimension $d$ with steady state  }

\label{sec_diff}

The inference for Langevin dynamics has been applied to many contexts
\cite{noise,mapsforce,biomolecule,singlemolecule,bacterial,decision,opticaltweezer,membrane,proteins,diffcoef,primerbayesian,metzler,cherstvy,movies,ronceray2020,underdamped,secondorder,learning,cellcell}.
In this section, the goal is to revisit this problem via the large deviations at Level 2.5 for diffusion processes
\cite{wynants_thesis,maes_diffusion,chetrite_formal,engel,chetrite_HDR,c_reset,c_lyapunov}.

\subsection{ Fokker-Planck generator parametrized by the force $\vec F(\vec x) $ and the diffusion coefficient $D(\vec x)$}

The Fokker-Planck dynamics
in the force field $\vec F(\vec x)$ with the diffusion coefficient $D(\vec x)$ in dimension $d$
\begin{eqnarray}
\frac{ \partial P_t(\vec x)   }{\partial t}   =  -   \vec \nabla .  \left[ P_t(\vec x )   \vec F(\vec x ) 
-D (\vec x) \vec \nabla   P_t(\vec x)  \right] \equiv {\cal F} P_t(.)
\label{fokkerplanck}
\end{eqnarray}
corresponds to a conserved continuity equation involving the probability density $P_t(\vec x) $ and the current
\begin{eqnarray}
\vec J_t(\vec x) \equiv P_t(\vec x )   \vec F(\vec x ) -D (\vec x) \vec \nabla   P_t(\vec x)  
\label{fokkerplanckj}
\end{eqnarray}
So here the Markov model $M$ that one wishes to infer
 is the Fokker-Planck generator $ {\cal F} $ parametrized by the force $\vec F(\vec x ) $ and the diffusion coefficient $ D(\vec x)$.

As in Eqs \ref{markovchainst} and \ref{mastereqst},
we will assume that the steady-state solution $P^*(\vec x)$ of Eq. \ref{fokkerplanck}
\begin{eqnarray}
0   =  {\cal F} P^*(.) =  -   \vec \nabla .  \left[ P^*(\vec x)  \vec F(\vec x ) 
-D (\vec x) \vec \nabla   P^*(\vec x)  \right]
\label{fokkerplanckst}
\end{eqnarray}
exists.
Again the steady state $P^*$ corresponds to the right eigenvector of the Fokker-Planck generator ${\cal F}$
associated to the eigenvalue zero, while the corresponding left eigenvector is constant.

\subsection{ Empirical observables and their constraints  }

For the Fokker-Planck dynamics of Eq. \ref{fokkerplanck}, 
the relevant time-empirical observables are :

(i) the empirical density
\begin{eqnarray}
 \rho(\vec x) && \equiv \frac{1}{T} \int_0^T dt \  \delta^{(d)} ( \vec x(t)- \vec x)  
\label{rhodiff}
\end{eqnarray}
normalized to unity
\begin{eqnarray}
\int d^d \vec x \ \rho (\vec x) && = 1
\label{rho1ptnormadiff}
\end{eqnarray}

(ii) the empirical current  
\begin{eqnarray} 
\vec j(\vec x) \equiv   \frac{1}{T} \int_0^T dt \ \frac{d \vec x(t)}{dt}   \delta^{(d)}( \vec x(t)- \vec x)  
\label{diffj}
\end{eqnarray}
that measures the time-average of the velocity $\frac{d \vec x(t)}{dt}$ 
 when the position $\vec x(t)$ at the same time $t$ is $\vec x$.
This empirical current should be divergence-free 
\begin{eqnarray}
 \vec \nabla . \vec j(\vec x) =0
\label{divergencenulle}
\end{eqnarray}
in order to be consistent with stationarity.


\subsection{ Large deviations at Level 2.5 for the empirical density $\rho(.)$ and the empirical current $\vec j(.)$ }

The joint probability distribution 
of the normalized empirical density $\rho(.)$ and of the empirical divergence-less current $\vec j(.)$
satisfies the large deviation form \cite{wynants_thesis,maes_diffusion,chetrite_formal,engel,chetrite_HDR,c_reset,c_lyapunov}
\begin{eqnarray}
 P_T[ \rho(.), \vec j(.)]   \opsimeq_{T \to +\infty}  \delta \left(\int d^d \vec x \rho(\vec x) -1  \right)
\left[ \prod_{\vec x }  \delta \left(  \vec \nabla . \vec j(\vec x) \right) \right] 
e^{- \displaystyle T I_{2.5} \left[ \rho(.);\vec j(.) \right]    }
\label{ld2.5diff}
\end{eqnarray}
where the rate function 
\begin{eqnarray}
  I_{2.5} \left[ \rho(.);\vec j(.) \right]   &&  =
  \int \frac{d^d \vec x}{ 4 D (\vec x) \rho(\vec x) } \left[ \vec j(\vec x) - \rho(\vec x) \vec F(\vec x)+D (\vec x) \vec \nabla \rho(\vec x) \right]^2
  \label{rate2.5diff}
\end{eqnarray}
vanishes only for the typical values corresponding to the steady-state of Eq. \ref{fokkerplanckst}
\begin{eqnarray}
  \rho^{typ}(\vec x)  && =P^*(\vec x) 
\nonumber \\
\vec j^{typ}(\vec x)   && =   P^*(\vec x)   \vec F(\vec x ) -D (\vec x) \vec \nabla   P^*(\vec x)  
\label{rhojtyp}
\end{eqnarray}

\subsection{ Inferred Fokker-Planck dynamics in the strict continuous-time limit  }

Reciprocally, from the empirical density $\rho(.)$ and the empirical current $\vec j(.)$ measured from the trajectory data,
the best inferred steady state $\hat P^*(.)$ and the best inferred force $\hat {\vec F}(.) $ are given by
\begin{eqnarray}
\hat   P^*(\vec x) && = \rho(\vec x)  
\nonumber \\
\hat {\vec F}(\vec x) && = \frac{ \vec j(\vec x) + D (\vec x) \vec \nabla   \rho (\vec x)  }{     \rho(\vec x)  }  
\label{FPinfer}
\end{eqnarray}
while the inferred diffusion coefficient $\hat D (\vec x) $ has to coincide with the true diffusion coefficient $D(\vec x)$
\begin{eqnarray} 
\hat D (\vec x) =    D (\vec x) 
\label{ddhat}
\end{eqnarray}
in the strict continuous-time limit, as explained in detail in Appendix \ref{app_pathintegral}
from the path-integral point of view.

\subsection{ Large deviations for the inferred Fokker-Planck parameters in the strict continuous-time limit  }

Via the change of variables of Eq \ref{FPinfer},
Eqs \ref{ld2.5diff} and \ref{rate2.5diff}
yields that the probability 
to infer the force $\hat {\vec F}(\vec x)  $ and the steady state $ \hat   P^*(.)$
of the corresponding Fokker-Planck generator $\hat {\cal F} $
\begin{eqnarray}
0= \hat {\cal F}  \hat P^*(\vec x) =  -   \vec \nabla .  \left[    \hat {\vec F}(\vec x )  \hat P^*(\vec x) -  D (\vec x) \vec \nabla   \hat P^*(\vec x)  \right]  
\label{fokkerplanckhatright}
\end{eqnarray}
follows the large deviation form
\begin{eqnarray}
 P^{Infer}_T[ \hat P^*(.), \hat {\vec F}(.) ] 
 \opsimeq_{T \to +\infty}  \delta \left(\int d^d \vec x \hat P^*(\vec x) -1  \right)
\left[ \prod_{\vec x }  \delta \left(  \vec \nabla .  \left[    \hat {\vec F}(\vec x )  \hat P^*(\vec x) - D (\vec x) \vec \nabla   \hat P^*(\vec x)  \right]  \right) \right] 
e^{- \displaystyle T {\cal I} \left[ \hat P^*(.), \hat {\vec F}(.)  \right]    }
\label{inferforcediff}
\end{eqnarray}
with the rate function 
\begin{eqnarray}
 {\cal I} \left[ \hat P^*(.), \hat {\vec F}(.) \right]  
 =
\int d^d \vec x   \hat   P^*(\vec x)
\left[    \frac{ \left(     \hat {\vec F}(\vec x) - \vec F(\vec x )  \right)^2
}{4 D(\vec x) }
 \right]
 \label{rateinferforcediff}
\end{eqnarray}
As in Eqs \ref{rateinferRing} and \ref{level2.5traprinfer},
the main qualitative conclusion is that the values $ \hat {\vec F}(\vec x) $ of the inferred force field 
are coupled via the inferred steady state $\hat P^*(.) $ that they produce together.

The impossibility to consider the fluctuations of the inferred diffusion coefficient (Eq. \ref{ddhat})
might be somewhat surprising, but it is due to the strict continuous-time limit as explained in detail in Appendix \ref{app_pathintegral} from the path-integral point of view.
However in practice, the numerical inference of diffusion processes is based on discretized data.
It is thus useful in the following subsections to re-analyze in detail the inference problem 
for diffusion processes from the point of view of discretized Langevin equations.


\subsection{ Equivalent Langevin dynamics with their discretized interpretations }

When the diffusion coefficient $D(\vec x)$ depends on the position $\vec x$,
the Fokker-Planck dynamics of Eq. \ref{fokkerplanck} corresponds to 
various Langevin stochastic differential equations involving $d$ independent Gaussian white noise components $\mu=1,..,d$
\begin{eqnarray}
\langle \eta_{\mu}(t) \rangle && =0
\nonumber \\
\langle \eta_{\mu}(t)\eta_{\nu}(t') \rangle && =\delta_{\mu,\nu} \delta(t-t')
 \label{whitenoise}
\end{eqnarray}
as follows.

\subsubsection{ Equivalent Langevin dynamics within the Stratonovich interpretation}

The Fokker-Planck dynamics of Eq. \ref{fokkerplanck} is equivalent to the Stratonovich Langevin dynamics
\begin{eqnarray}
\frac{d \vec x (t)}{dt} = \vec f_{S}(\vec x(t)) + \sqrt{ 2 D (\vec x(t) ) } \  \vec \eta(t) 
 \label{strato}
\end{eqnarray}
with the effective Stratonovich force
\begin{eqnarray}
\vec f_{S}(\vec x) \equiv \vec F(\vec x ) +\frac{1}{2}  \vec \nabla D (\vec x )
 \label{fstrato}
\end{eqnarray}
Eq. \ref{strato}
 should be interpreted with the mid-point discretization scheme for the multiplicative factor of the noise
\begin{eqnarray}
\vec x(t+\Delta t)  = \vec x(t) + \vec f_{I}(\vec x(t)) \Delta t + \sqrt{  D (\vec x(t+\Delta t) ) +  D (\vec x(t) )} \int_t^{t+\Delta t} dt' \vec \eta(t') 
 \label{stratodiscrete}
\end{eqnarray}

\subsubsection{ Equivalent Langevin dynamics within the Ito interpretation}

The Fokker-Planck dynamics of Eq. \ref{fokkerplanck} is equivalent to the Ito Langevin dynamics
\begin{eqnarray}
\frac{d \vec x (t)}{dt} = \vec f_{I}(\vec x(t)) + \sqrt{ 2 D (\vec x(t) ) } \  \vec \eta(t) 
 \label{ito}
\end{eqnarray}
with the effective Ito force
\begin{eqnarray}
\vec f_{I}(\vec x) \equiv \vec F(\vec x ) + \vec \nabla D (\vec x )
 \label{fito}
\end{eqnarray}
Eq. \ref{ito}
 should be interpreted with the causal discretization scheme
\begin{eqnarray}
\vec x(t+\Delta t)  = \vec x(t) + \vec f_{I}(\vec x(t)) \Delta t + \sqrt{ 2 D (\vec x(t) ) } \int_t^{t+\Delta t} dt' \vec \eta(t') 
 \label{itodiscrete}
\end{eqnarray}
i.e. the corresponding Gaussian propagator for the time-interval $\Delta t$ reads
\begin{eqnarray}
W^{[\Delta t]} ( \vec x(t+\Delta t) \vert \vec x(t) ) && = 
\left( \frac{1}{  4 \pi  D (\vec x(t) ) \Delta t}  \right)^{\frac{d}{2} }
e^{ \displaystyle  - \frac{ \left( \vec x(t+\Delta t)  - \vec x(t) - \vec f_{I}(\vec x(t)) \Delta t \right)^2 }
{ 4 D (\vec x(t) ) \Delta t}  }
\nonumber \\
&& = 
\left( \frac{1}{  4 \pi  D (\vec x(t) ) \Delta t}  \right)^{\frac{d}{2} }
e^{ \displaystyle  - \Delta t \frac{ \left( \frac{\vec x(t+\Delta t)  - \vec x(t)}{ \Delta t }  - \vec f_{I}(\vec x(t)) \right)^2 }
{ 4 D (\vec x(t) )   } }
 \label{itodiscretegaussian}
\end{eqnarray}


\subsection{ Inference for the Markov chain kernel corresponding to the Ito discretization of the Langevin equation}

\label{subsec_ito}

Eq. \ref{itodiscretegaussian} corresponds to the discrete-time continuous-space Markov chain Gaussian kernel 
of parameters $[\vec f_{I}(.);D(.)]$
\begin{eqnarray}
W^{[\Delta t]}_{[\vec f_{I}(.);D(.)]} ( \vec x \vert \vec y ) && = 
\left( \frac{1}{  4 \pi  D (\vec y ) \Delta t}  \right)^{\frac{d}{2} }
e^{ \displaystyle 
 - \frac{ \left( \vec x  - \vec y - \vec f_{I}(\vec y) \Delta t \right)^2 }{ 4 D (\vec y ) \Delta t} 
 }
 \label{chaingaussian}
\end{eqnarray}
while the time $T$ corresponds to $N= \frac{T}{\Delta t}$ time steps.
As a consequence, 
Eq. \ref{proba2.5chaininfercontinuous} yields that the probability to
 infer the generator $W^{[\Delta t]}_{[ {\hat {\vec f}}_I(\vec .) , \hat D (. )  ]} $
together with its steady state $\hat P^*(.)  $ 
follows the large deviation form
\begin{eqnarray}
&& P_T^{Infer} ( W_{[ {\hat {\vec f}}_I( .) , \hat D (. )  ]}(.,.)  ; \hat P^*(.) )
  \opsimeq_{T \to +\infty}  
  \delta \left( \int d^d \vec x \  \hat P^*(\vec x) - 1 \right) 
 \prod_{\vec x}  \delta \left(    \int d^d \vec y \  
 W_{[ {\hat {\vec f}}_I( .) , \hat D (. )  ]} (\vec x, \vec y)  \hat P^*( \vec y) - \hat P^*( \vec x)\right)
 \nonumber \\
&&  e^{ \displaystyle - \frac{T}{\Delta t}  \int d^d \vec y  \ \hat P^*( \vec y)
  \int d^d \vec x \    W_{[ {\hat {\vec f}}_I( .) , \hat D (. )  ]} (\vec x, \vec y) 
  \ln \left( \frac{ W_{[ {\hat {\vec f}}_I(\ .) , \hat D (. )  ]}(\vec x, \vec y) }
  {  W_{[\vec f_I(.), D(.)] }   (\vec x, \vec y)  }  \right)  } 
\label{probainfergauss}
\end{eqnarray}
where the last constraint on the first line of Eq. \ref{proba2.5chaininfercontinuous} does not appear,
since the conservation of probability is always satisfied by the generator $W^{[\Delta t]}_{[ {\hat {\vec f}}_I(\vec .) , \hat D (. )  ]} $ 
(see the form of Eq. \ref{chaingaussian}).

The last integral of the rate function  in the exponential reads using the Gaussian kernel of Eq. \ref{chaingaussian}
\begin{eqnarray}
&&  
 \int d^d \vec x \    W_{[ {\hat {\vec f}}_I( .) , \hat D (. )  ]} (\vec x, \vec y) 
  \ln \left( \frac{ W_{[ {\hat {\vec f}}_I(\ .) , \hat D (. )  ]}(\vec x, \vec y) }
  {  W_{[\vec f_I(.), D(.)] }   (\vec x, \vec y)  }  \right) 
  \nonumber \\
&& =  \int d^d \vec x \   
\left( \frac{1}{  4 \pi  \hat D (\vec y ) \Delta t}  \right)^{\frac{d}{2} }
e^{ \displaystyle  - \frac{ \left( \vec x  - \vec y - {\hat {\vec f}}_{I}(\vec y) \Delta t \right)^2 }
{ 4 \hat D (\vec y ) \Delta t}  }  
\left[ \frac{d}{2} \ln \left( \frac{  D (\vec y )}{ \hat D (\vec y )   }  \right) 
+ \frac{ \left( \vec x  - \vec y - \vec f_{I}(\vec y) \Delta t \right)^2 }{ 4 D (\vec y ) \Delta t  }
- \frac{\left( \vec x  - \vec y - {\hat {\vec f}}_{I}(\vec y) \Delta t \right)^2 }{ 4 \hat D (\vec y ) \Delta t }
\right]
  \nonumber \\
&& = \frac{ \left[  {\hat {\vec f}}_I(\vec y) -  \vec f_I(\vec y) \right]^2 }{ 4 D (\vec y ) } \Delta t
+ \frac{ d  }{ 2 } \left[ - \ln \left( \frac{  \hat D (\vec y )}{  D (\vec y )   }  \right) 
+  \frac{\hat D (\vec y )  }{  D (\vec y )}-1 \right]
\label{entropyrelativegauss}
\end{eqnarray}
so that Eq. \ref{probainfergauss}
reduces to
\begin{eqnarray}
&& P_T^{Infer} ( W_{[ {\hat {\vec f}}_I( .) , \hat D (. )  ]}(.,.)  ; \hat P^*(.) )
  \opsimeq_{T \to +\infty}  
 \nonumber \\
&&   \delta \left( \int d^d \vec x \  \hat P^*(\vec x) - 1 \right) 
 \prod_{\vec x}  \delta \left(    \int d^d \vec y \  
 \left( \frac{1}{  4 \pi  \hat D (\vec y ) \Delta t}  \right)^{\frac{d}{2} }
e^{   - \frac{ \left( \vec x  - \vec y - {\hat {\vec f}}_{I}(\vec y) \Delta t \right)^2 }
{ 4 \hat D (\vec y ) \Delta t}  }  
   \hat P^*( \vec y) - \hat P^*( \vec x)\right)
 \nonumber \\
&&  e^{ \displaystyle - T  \int d^d \vec y  \ \hat P^*( \vec y)
 \left( 
 \frac{ \left[  {\hat {\vec f}}_I(\vec y) -  \vec f_I(\vec y) \right]^2 }{ 4 D (\vec y ) } 
+ \frac{ d  }{ 2 \Delta t} \left[ - \ln \left( \frac{  \hat D (\vec y )}{  D (\vec y )   }  \right) 
+  \frac{\hat D (\vec y )  }{  D (\vec y )}-1 \right]
 \right)
 }
\label{probainfergaussito}
\end{eqnarray}

In conclusion, it is important to distinguish whether the time step $\Delta t$ remains finite or tends towards zero :

(i) if the time-step $\Delta t$ of the Ito discretization scheme of Eq. \ref{itodiscrete}
remains finite, then the probability to infer the Ito force $\vec f_I(.) $ and the diffusion coefficient $\hat D (. )$
follows the large deviation form of Eq. \ref{probainfergaussito}.

(ii) in the limit $\Delta t \to 0$, the last term of the rate function of Eq. \ref{probainfergaussito} diverges if $\hat D(\vec y ) \ne D(\vec y) $.
One recovers that the inferred diffusion coefficient $\hat D (\vec x) $ has to coincide with the true diffusion coefficient $D(\vec x)$
as in Eq. \ref{ddhat}.
Within the present Ito discretization computation, the origin of this property is 
the divergence of the number $N=\frac{T}{\Delta t} \to +\infty$ of time-steps in the limit $\Delta t \to 0 $,
so that the inferred diffusion $\hat D (\vec x) $ coefficient cannot fluctuate any more but is fixed to its typical value given by the true diffusion coefficient.


\section{ Application to the diffusion in a disordered potential in dimension $d$  }

\label{sec_diffpotential}

In this section, the large deviations analysis of inference for diffusion processes of the previous section
is applied to the thermal equilibrium diffusion in a disordered potential $U(\vec x)$ in dimension $d$.

\subsection{ Model and steady state  }

In this section, we consider the Fokker-Planck dynamics of Eq. \ref{fokkerplanck}
for the special case 
where the diffusion coefficient is uniform and fixed by the inverse temperature $\beta$
(note that in the present paper, the notation $T$ represents the time window of the trajectory and not the temperature,
so that the temperature will only appear via its inverse $\beta$)
\begin{eqnarray}
D (\vec x) = \frac{1}{\beta }
\label{Dinversetemp}
\end{eqnarray}
while the force $\vec F(\vec x )  $ derives from some disordered potential $U(\vec x)$
\begin{eqnarray}
\vec F (\vec x) =  - \vec \nabla U(\vec x)
\label{FderiU}
\end{eqnarray}
So the Fokker-Planck dynamics of Eq. \ref{fokkerplanck} becomes
\begin{eqnarray}
\frac{ \partial P_t(\vec x)   }{\partial t}   =  -   \vec \nabla .  \left[ P_t(\vec x )  \left( - \vec \nabla U(\vec x) \right) 
-  \frac{1}{\beta } \vec \nabla   P_t(\vec x)  \right] 
\label{fokkerplanckexample}
\end{eqnarray}
The steady state of Eq. \ref{fokkerplanckst} corresponds to the Boltzmann equilibrium
at inverse temperature $\beta$
in the potential $U(\vec x) $ on the appropriate domain $\vec x \in V$ for the model under study
\begin{eqnarray}
 P^*(\vec x)  = \frac{ e^{ - \beta  U(\vec x)} }{ \int_V d^d \vec x'  e^{ - \beta  U(\vec x' )} }
\label{equil}
\end{eqnarray}
where the partition function of the denominator ensures the normalization of the steady state over the domain $V$.

\subsection{ Large deviations for the inference of the disordered potential $U(\vec x)$  }

Eqs \ref{inferforcediff}
and \ref{rateinferforcediff}
yields that the probability to infer the potential $\hat U(\vec x)$
instead of the true potential $U(\vec x)$ 
follows the large deviation form with respect to the time-window $T$ of the trajectory 
\begin{eqnarray}
 P^{Infer}_T[ \hat U(.) ] 
 \opsimeq_{T \to +\infty}  e^{- \displaystyle T {\cal I} \left[ \hat U(.)  \right]    }
\label{inferforcediffu}
\end{eqnarray}
with the explicit rate function 
\begin{eqnarray}
 {\cal I} \left[ \hat U(.) \right]  
 = \left(\frac{\beta}{4} \right)
  \frac{ \displaystyle \int_V d^d \vec x  \left(       \vec \nabla \hat U(\vec x) - \vec \nabla U(\vec x)  \right)^2  e^{ - \beta  \hat U(\vec x)} }
  {\displaystyle \int_V d^d \vec x'  e^{ - \beta \hat U(\vec x' )} }
 \label{rateinferforcediffu}
\end{eqnarray}
As in the other previous examples (see Eqs \ref{rateinferRing},\ref{level2.5traprinfer} and \ref{rateinferforcediff}),
the values $\hat U(\vec x) $ of the inferred potential are coupled 
via the inferred steady state $\hat P^*(.) $ that they produce together,
given here by the corresponding Boltzmann equilibrium 
\begin{eqnarray}
 {\hat P}^*(\vec x)  = \frac{ e^{ - \beta  \hat U(\vec x)} }{ \int_V d^d \vec x'  e^{ - \beta  \hat U(\vec x' )} }
\label{equilhat}
\end{eqnarray}


\section{ Conclusion  }

\label{sec_conclusion}

In this paper, we have revisited the inference of Markov models via the large deviations properties at Level 2.5
within the following point of view : 
\\ ~ \\
(1) The input is a single dynamical trajectory $x(0 \leq t \leq T)$ over a very long time $T$ 
\\ ~ \\
(2) From the trajectory data of (1), one computes the following empirical time-averaged observables (that should not depend too much on the time window $T$, otherwise this means that $T$ is not large enough to see the convergence towards some steady state) :

(2a) the empirical time-averaged density $\rho_x$, given by Eq \ref{rho1pt} in discrete-time, or by Eq. \ref{rho1pj} in continuous-time

(2b)  the empirical density of jumps, given by $\rho^{(2)}_{x,y}$ of Eq \ref{rho2pt} in discrete-time, or by $q_{x,y}$ of Eq \ref{jumpempiricaldensity} in continuous-time 
\\ ~ \\
(3) The best Markov model that can be inferred from the empirical time-averaged observables of (2) 
via the principle of maximum likelihood is as follows :

(3a) the best inferred steady-state $\hat P^*$ is directly the empirical density $\rho_x$.
Note that the possible states in the inferred model are given by the different states 
that are actually seen in the empirical density of the data in (2a).
If some states of the true model have a too small probability in the steady state to be visited during the trajectory $x(0 \leq t \leq T)$,
these states will not appear in the inferred model.

(3b) the best inferred Markov matrix, given by $\hat W_{x ,y} = \frac{\rho^{(2)}_{x,y}}{ \rho_{y}}  $ of Eq. \ref{Winfer}
in discrete-time, or by $\hat w_{x ,y} \equiv \frac{q_{x,y}}{ \rho_{y}}$ of Eq. \ref{winfer} in continuous-time. 
Note that the possible jumps between states in the inferred model are given by the different jumps
that are actually seen in the empirical jump-density of the data in (2b). If some possible jumps of the true model
 have a too small probability in the steady state to be visited during the trajectory $x(0 \leq t \leq T)$,
these jumps will not appear in the inferred model.
 \\ ~ \\
4) The fluctuations of the best inferred Markov matrix of (3b) and of its corresponding steady state of (3a)
with respect to the 'true' Markov matrix and the true steady state is given by the large deviations of Eq. \ref{proba2.5chaininfer}
in discrete-time, or of Eq. \ref{level2.5masterinfer} in continuous-time. \\ ~ \\ 

Applications to various models of random walks in disordered media have been described, 
where the goal was to infer the quenched disordered variables defining a given disordered sample.
Given the recent availability of big data in many fields and the corresponding extensive use of inference methods,
we hope that the present analysis can be useful to characterize 
 the statistical fluctuations of the inferred parameters of Markov models
with respect to the 'true' values of these parameters.

As a final remark, we should stress that for simplicity, we have only considered 
the case of time-independent Markov models with steady-states.
However the Level 2.5 has been also formulated for other frameworks, 
so let us mention two possible natural extensions of the present framework :

(i) when the dynamics does not converge towards some steady state,
 or when the Markov matrix is time-dependent,
one should use instead the formulation of Ref. \cite{maes_onandbeyond}, where the Level 2.5 concerns
the ensemble-averaged observables at fixed time, i.e. one should consider that the input in (1)
is not a single trajectory $x(0 \leq t \leq T)$ over a very long time $T$ anymore,
but instead a large number $N$ of independent trajectories $x_n(0 \leq t \leq T)$ labelled by the index $n=1,2,..,N$
over the finite time $T$  that one wishes to analyze for the inference problem.

(ii) when the Markov matrix is time-periodic, one should instead use the formulation of Ref. \cite{barato_periodic,chetrite_periodic},
where the Level 2.5 concerns the empirical period-averaged observables for any given position inside the period.


\appendix

\section{ Inference for diffusion processes via the path-integral approach }

\label{app_pathintegral}

Since the property of Eq. \ref{ddhat} concerning the inference of the diffusion coefficient for the Fokker-Planck dynamics
can be very surprising at first,
it is useful in this Appendix to discuss in details the origin of this property from 
the point of view of the general principles of section \ref{sec_general} using path-integral methods.


\subsection{Identification of the relevant time-empirical observables that determine the trajectories probabilities } 

For the Fokker-Planck dynamics of Eq. \ref{fokkerplanck},
the probability of the trajectory $x(0 \leq t \leq T)$ 
\begin{eqnarray}
{\cal P}[\vec x(0 \leq t \leq T)] = e^{ - \displaystyle 
  \int_0^{T} dt \left[  \frac{[ \frac{d \vec x (t)}{dt} - \vec F(\vec x (t)) ]^2 }{4D(\vec x(t))}
 - \frac{  [\vec \nabla D(\vec x(t)) ]^2}{16D(\vec x(t))} + \frac{ \Delta D(\vec x(t)) }{4} + \frac{ \vec \nabla . \vec F(\vec x(t)) }{2}
\right]
 } \prod_{t} \left( D( \vec x(t) )\right)^{- \frac{ d}{2} }
\label{pathFP}
\end{eqnarray}
involves both the action in the exponential and the non-trivial measure-factor
that depends on the diffusion coefficient $D( \vec x(t)) $ along the trajectory.
If one wishes to include this measure-factor in the action, 
one needs to introduce some regularization with some time-step $\Delta t$
and the $N= \frac{T}{\Delta t}$ times $t_k=k \Delta t$ with $k=1,2,..,N$ to obtain
\begin{eqnarray}
\left[ \prod_{t} \left( D( \vec x(t) )\right)^{- \frac{ d}{2} } \right]^{Reg} 
= \prod_{k=1}^{\frac{T}{\Delta t}} \left[ D( \vec x(t= k \Delta t) )\right]^{- \frac{ d}{2} }
=  e^{ \displaystyle - \frac{ d}{2} \sum_{k=1}^{\frac{T}{\Delta t}} \ln \left[ D( \vec x(t= k \Delta t) ) \right] }
\simeq e^{ \displaystyle - \frac{ d}{2 \Delta t} \int_0^T dt \ln \left[ D( \vec x(t) ) \right] }
\label{regulari}
\end{eqnarray}

In the trajectory probability of Eq. \ref{pathFP},
the relevant time-empirical observables that appear are thus not only
the empirical density of Eq. \ref{rhodiff} and the empirical current of Eq. \ref{diffj} discussed in the main text,
but also the empirical kinetic energy 
\begin{eqnarray} 
k(\vec x) && \equiv  
 \frac{1}{T} \int_0^T dt \ \frac{1}{2} \left(\frac{d \vec x(t)}{dt} \right)^2  \delta^{(d)}( \vec x(t)- \vec x)  
 \label{kineticempi}
\end{eqnarray}
that measures the time-average of the kinetic energy $\frac{1}{2} \left(\frac{d \vec x(t)}{dt} \right)^2 $
 when the position $\vec x(t)$ at the same time $t$ is $\vec x$.

If one introduces the following notation for the normalized average of an arbitrary observable ${\cal O}$ at position $\vec x$
\begin{eqnarray} 
\langle {\cal O}\rangle_{\vec x}  && \equiv \frac{  \frac{1}{T} \int_0^T dt \ {\cal O}  \delta^{(d)}( \vec x(t)- \vec x)  }
{  \frac{1}{T} \int_0^T dt \   \delta^{(d)}( \vec x(t)- \vec x)  }
= \frac{  \frac{1}{T} \int_0^T dt \ {\cal O}  \delta^{(d)}( \vec x(t)- \vec x)  }
{ \rho (\vec x)   }
 \label{average}
\end{eqnarray}
one obtains in terms of the empirical observables 
\begin{eqnarray} 
\langle 1 \rangle_{\vec x}  && =1
\nonumber \\
\langle \frac{d \vec x(t)}{dt}\rangle_{\vec x}  && = \frac{ \vec j(\vec x)  } { \rho (\vec x)   }
\nonumber \\
\langle \frac{1}{2} \left(\frac{d \vec x(t)}{dt} \right)^2 \rangle_{\vec x}  && = \frac{  k(\vec x)  } { \rho (\vec x)   }
 \label{average3}
\end{eqnarray}
As a consequence, the positivity of the variance of the velocity gives the following constraint for the 
empirical kinetic energy $k(\vec x)$
\begin{eqnarray} 
 \langle \left(\frac{d \vec x(t)}{dt} - \langle\frac{d \vec x(t)}{dt} \rangle_{\vec x}\right)^2 \rangle_{\vec x}  
 =  \langle \left(\frac{d \vec x(t)}{dt} \right)^2 \rangle_{\vec x} 
 -   \left( \langle\frac{d \vec x(t)}{dt} \rangle_{\vec x}\right)^2 = 
  \frac{ 2 k(\vec x)  } { \rho (\vec x)   } - \left(  \frac{ \vec j(\vec x)  } { \rho (\vec x)   }\right)^2  
  =  \frac{ 2   } { \rho (\vec x)   } \left[k(\vec x) -   \frac{ \vec j^2(\vec x)  } { 2 \rho (\vec x)  } \right] \geq 0
 \label{kconstraint}
\end{eqnarray}
It is thus more convenient to replace the empirical kinetic energy $k(\vec x)  $
by the empirical positive excess of kinetic energy
\begin{eqnarray} 
 e(\vec x) \equiv    \frac{ k(\vec x)   } { \rho (\vec x)   }  -  \frac{1}{2} \left( \frac{  \vec j(\vec x)  } {  \rho (\vec x)  } \right)^2 \geq 0
 \label{excessdef}
\end{eqnarray}
i.e. the empirical kinetic energy $k(\vec x)  $ is decomposed into the two contributions
\begin{eqnarray} 
   k(\vec x)      =    \frac{  \vec j^2(\vec x)  } { 2 \rho (\vec x)  } +  \rho (\vec x) e(\vec x)
 \label{kreplace}
\end{eqnarray}
In terms of these empirical observables,
the probability of the trajectory $x(0 \leq t \leq T)$ of Eq. \ref{pathFP} 
with the regularized form of Eq. \ref{regulari} 
can be rewritten as
\begin{eqnarray}
&& {\cal P}[\vec x(0 \leq t \leq T)] 
\nonumber \\
&& = e^{ - \displaystyle 
 T \int d^d \vec x \left(  \frac{ k(\vec x) }{2D(\vec x)}
  -  \frac{ \vec j(\vec x) . \vec F(\vec x )  }{2D(\vec x)}
  +  \rho(\vec x) 
\left[    \frac{[  \vec F(\vec x ) ]^2 }{4D(\vec x)}
 - \frac{  [\vec \nabla D(\vec x) ]^2}{16D(\vec x)} + \frac{ \Delta D(\vec x) }{4} + \frac{ \vec \nabla . \vec F(\vec x) }{2}
 \right]
\right)
 } e^{ \displaystyle - \frac{ d T }{2 \Delta t} \int d^d \vec x \rho(\vec x) \ln \left[ D( \vec x ) \right] }
\nonumber \\
&& = e^{ - \displaystyle 
 T \int d^d \vec x  \rho(\vec x) 
\left[    \frac{ \left(  \frac{\vec j(\vec x) }{  \rho(\vec x) } - \vec F(\vec x )\right)^2 }{4 D(\vec x) }
+  \frac{   e(\vec x) }{2D(\vec x)}
 - \frac{  [\vec \nabla D(\vec x) ]^2}{16D(\vec x)} + \frac{ \Delta D(\vec x) }{4} + \frac{ \vec \nabla . \vec F(\vec x) }{2}
  + \frac{ d  }{2 \Delta t}  \ln \left[ D( \vec x ) \right]
 \right]
 }
\label{pathFPempi}
\end{eqnarray}

With respect to the general formalism summarized in Section \ref{sec_general},
this means that the relevant empirical observables $ E$ are the empirical density $\rho(.)$, the empirical current $\vec j(.) $
and the empirical excess of kinetic energy $e(.)$,
while the intensive action introduced in Eq. \ref{ptrajectempi} reads
\begin{eqnarray}
&& A_{\vec F(.);D(.)} \left( [\rho(.);\vec j(.); e(.)] \right) =
\nonumber \\
&& \int d^d \vec x  \rho(\vec x) 
\left[    \frac{ \left(  \frac{\vec j(\vec x) }{  \rho(\vec x) } - \vec F(\vec x )\right)^2 }{4 D(\vec x) }
+  \frac{   e(\vec x) }{2D(\vec x)}
 - \frac{  [\vec \nabla D(\vec x) ]^2}{16D(\vec x)} 
 + \frac{ \Delta D(\vec x) }{4} 
 -  \frac{ \vec F(\vec x) . \vec \nabla \rho(\vec x) }{2 \rho(\vec x)}
+ \frac{ d  }{2 \Delta t}  \ln \left[ D( \vec x ) \right] \right]
\label{actiondiff}
\end{eqnarray}
The penultimate term has been rewritten via an integration by parts with respect to Eq. \ref{pathFPempi}.


\subsection{ Typical values of the empirical observables }

The typical values for the empirical density and for the current have been given in Eq. \ref{rhojtyp}
of the text,
while the typical value of the empirical excess of kinetic energy $e(.)$
 is given by the diffusion coefficient
\begin{eqnarray} 
e^{typ}(\vec x) =D(x) 
 \label{etyp}
\end{eqnarray}

Reciprocally, the modified Fokker-Planck operator $\hat {\cal F}$ 
that would make these three empirical observables $[\rho(.);\vec j(.); e(.)]$ typical
is parametrized by the
modified diffusion coefficient 
\begin{eqnarray} 
 \hat D (\vec x)  =  e(\vec x)
\label{diffinfer}
\end{eqnarray}
and by the modified force 
\begin{eqnarray}
\hat {\vec F}(\vec x) = \frac{ \vec j(\vec x) + e (\vec x) \vec \nabla   \rho (\vec x)  }{     \rho(\vec x)  }  
\label{forceinfer}
\end{eqnarray}

With respect to the general formalism summarized in Section \ref{sec_general},
this means that the intensive action of Eq. \ref{actiondiff} reads for the modified 
Fokker-Planck operator $\hat {\cal F}$ with the parameters given by Eqs \ref{diffinfer} and \ref{forceinfer}
\begin{eqnarray}
&& A_{\hat {\vec F}(.); \hat D(.)} \left( [\rho(.);\vec j(.); e(.)] \right) 
 \nonumber \\
 && =
 \int d^d \vec x  \rho(\vec x) 
\left[    \frac{ \left(  \frac{\vec j(\vec x) }{  \rho(\vec x) } - \hat {\vec F}(\vec x )\right)^2 }{4 \hat D(\vec x) }
+  \frac{   e(\vec x) }{2 \hat D(\vec x)}
 - \frac{  [\vec \nabla \hat D(\vec x) ]^2}{16 \hat D(\vec x)} 
 + \frac{ \Delta \hat D(\vec x) }{4} 
 -  \frac{ \hat {\vec F}(\vec x) . \vec \nabla \rho(\vec x) }{2 \rho(\vec x)}
 + \frac{ d  }{2 \Delta t}  \ln \left[ \hat D( \vec x ) \right]
 \right]
 \nonumber \\
&&  =
 \int d^d \vec x  \rho(\vec x) 
\left[    \frac{e (\vec x) \left(  \frac{   \vec \nabla   \rho (\vec x)  }{     \rho(\vec x)  }  \right)^2 }{4  }
+  \frac{   1 }{2 }
 - \frac{  [\vec \nabla e(\vec x) ]^2}{16 e(\vec x)} 
 + \frac{ \Delta e(\vec x) }{4}  
  -  \frac{ \left( \frac{ \vec j(\vec x) + e (\vec x) \vec \nabla   \rho (\vec x)  }{     \rho(\vec x)  }  \right) . \vec \nabla \rho(\vec x) }{2 \rho(\vec x)}
  + \frac{ d  }{2 \Delta t}  \ln \left[ e( \vec x ) \right]
 \right]
 \label{actiondifftyp}
\end{eqnarray}


\subsection{ Large deviations for the relevant time-empirical observables }

So the joint distribution of the empirical density $\rho(.)$,the empirical current $\vec j(.)$
and the empirical excess of kinetic energy $e(.)$
satisfy the large deviation form 
\begin{eqnarray}
 P_T[ \rho(.), \vec j(.); e(.)]   \opsimeq_{T \to +\infty}  \delta \left(\int d^d \vec x \rho(\vec x) -1  \right)
\left[ \prod_{\vec x }  \delta \left(  \vec \nabla . \vec j(\vec x) \right) \right] 
e^{- \displaystyle T I_{2.75} \left[ \rho(.);\vec j(.); e(.) \right]    }
\label{ld2.75diff}
\end{eqnarray}
where the rate function $ I_{2.75} \left[ \rho(.);\vec j(.); e(.) \right] $ of Level 2.75
(this name just means that it is higher than the usual 2.5 Level discussed below in Eq. \ref{rate2.5diff2.75})
corresponds to the difference (Eq. \ref{rateempi}) 
between the actions of Eqs \ref{actiondiff} and \ref{actiondifftyp}
\begin{eqnarray}
&& I_{2.75} \left[ \rho(.);\vec j(.); e(.) \right]  
= A_{\vec F(.);D(.)} \left( [\rho(.);\vec j(.); e(.)] \right)  
- A_{\hat {\vec F}(.); \hat D(.)} \left( [\rho(.);\vec j(.); e(.)] \right) 
\nonumber \\
&& =\int d^d \vec x  \rho(\vec x) 
\left[    \frac{ \left(  \frac{\vec j(\vec x) }{  \rho(\vec x) } - \vec F(\vec x )\right)^2 }{4 D(\vec x) }
+  \frac{   e(\vec x) }{2D(\vec x)}
 - \frac{  [\vec \nabla D(\vec x) ]^2}{16D(\vec x)} 
 + \frac{ \Delta D(\vec x) }{4} 
 -  \frac{ \vec F(\vec x) . \vec \nabla \rho(\vec x) }{2 \rho(\vec x)} \right] 
\nonumber \\
&&
- \int d^d \vec x  \rho(\vec x) 
\left[    \frac{e (\vec x) \left(  \frac{   \vec \nabla   \rho (\vec x)  }{     \rho(\vec x)  }  \right)^2 }{4  }
+  \frac{   1 }{2 }
 - \frac{  [\vec \nabla e(\vec x) ]^2}{16 e(\vec x)} 
 + \frac{ \Delta e(\vec x) }{4}  
  -  \frac{ \left( \frac{ \vec j(\vec x) + e (\vec x) \vec \nabla   \rho (\vec x)  }{     \rho(\vec x)  }  \right) . \vec \nabla \rho(\vec x) }{2 \rho(\vec x)} \right]
 \nonumber \\
&& +\frac{ d  }{2 \Delta t}   \int d^d \vec x  \rho(\vec x)   \ln \left[ \frac{ D( \vec x ) }{e( \vec x ) } \right]
\label{lrated2.75diff}
\end{eqnarray}
where we have written separately on the last line the regularized contribution 
involving the time-step $\Delta t$ (see Eq. \ref{regulari}).
In the limit $\Delta t \to 0$, this contribution becomes singular unless the
empirical excess of kinetic energy $e(.)$ coincides with the diffusion coefficient $D(.)$
\begin{eqnarray} 
   e(\vec x) = D (\vec x) 
\label{edcoincide}
\end{eqnarray}
This means that the empirical excess of kinetic energy $e(.)$ is actually not allowed to fluctuate
but is fixed by Eq. \ref{edcoincide} to its typical value.

As a consequence, in the strict continuous-time limit $\Delta t \to 0$,
the only empirical observables that can fluctuate are
the empirical density $\rho(.)$
and the empirical current $\vec j(.)$ as mentioned in the text,
and their large deviations properties are described by Eq. \ref{ld2.5diff},
where the rate function of Eq. \ref{rate2.5diff}
is given by Eq. \ref{ld2.75diff}
for the case $e(\vec x) = D (\vec x) $ of Eq. \ref{edcoincide}
\begin{eqnarray}
&&  I_{2.5} \left[ \rho(.);\vec j(.) \right]   =  I_{2.75} \left[ \rho(.);\vec j(.); e(.)=D(.) \right]  
\nonumber \\
&& = \int d^d \vec x  \rho(\vec x) 
\left[    \frac{ \left( \vec F(\vec x ) - \frac{\vec j(\vec x) }{  \rho(\vec x) } \right)^2 - D^2 (\vec x) \left(  \frac{   \vec \nabla   \rho (\vec x)  }{     \rho(\vec x)  }  \right)^2}{4 D(\vec x) }
 -  \frac{ \left( \vec F(\vec x) - \frac{ \vec j(\vec x) + D (\vec x) \vec \nabla   \rho (\vec x)  }{     \rho(\vec x)  } \right) . \vec \nabla \rho(\vec x) }{2 \rho(\vec x)} \right] 
\nonumber \\
&&  =\int d^d \vec x  \rho(\vec x) 
\left[    \frac{ 
 \left(\vec F(\vec x ) -  \frac{\vec j(\vec x) + D (\vec x)  \vec \nabla   \rho (\vec x)}{  \rho(\vec x) } \right)^2
}{4 D(\vec x) }
 \right]
\label{rate2.5diff2.75}
\end{eqnarray}


\subsection{ Consequence for the inferred diffusion coefficient  }

So the inferred diffusion coefficient $\hat D (\vec x) $ has to coincide with the true diffusion coefficient $D(\vec x)$
in the strict continuous-time limit $\Delta t \to 0$ (see Eqs \ref{diffinfer} and \ref{edcoincide})
\begin{eqnarray} 
\hat D (\vec x) =   e(\vec x) = D (\vec x) 
\label{edcoincidehat}
\end{eqnarray}
Within the present path-integral approach, the origin of this property is 
the non-trivial measure-factor of Eq. \ref{regulari}
in the trajectory probability of Eq. \ref{pathFP}
that would give an infinite result in the last contribution of the rate function of Eq. \ref{lrated2.75diff}
 for any other diffusion coefficient $ \hat D (\vec x) \ne D (\vec x) $ than the true diffusion coefficient $D (\vec x) $.
Another perspective of the property of Eq. \ref{edcoincidehat}
is given in the text after Eq. \ref{probainfergaussito}.



\begin{thebibliography}{99}


\bibitem{book_information}
T. M. Cover and J. A. Thomas, Elements of Information Theory, J. Wiley, New Jersey (2006).
  
\bibitem{book_inference}
 L. Wasserman, All of statistics: a concise course in statistical inference, Springer  (2020)

\bibitem{book_learning}
 D. J. C. MacKay, Information theory, inference and learning algorithms, Cambridge University Press (2003)



\bibitem{bayesmarkov}
A. Gomez‐Corral,  D. Rios Insua,  F. Ruggeri and  M. Wiper,
Wiley Statistics Reference Online 2015 
https://doi.org/10.1002/9781118445112.stat07837


\bibitem{1957}
T. W. Anderson and L. A. Goodman,
The Annals of Mathematical Statistics, 28, 89 (1957).

\bibitem{1961}
P. Billingsley, Ann. Math. Statist. 32, 12 (1961).

\bibitem{zeitouni}
A. Dembo and O. Zeitouni,
Large Deviations Techniques and Applications, Springer Berlin (2010).



\bibitem{evolution}
M. A. Suchard, R. E. Weiss, and J. S. Sinsheimer,
Mol. Biol. Evol. 18, 1001 (2001).

\bibitem{monassonDNAprl}
V. Baldazzi, S. Cocco, E. Marinari, and R. Monasson,
Phys. Rev. Lett. 96, 128102 (2006).

\bibitem{monassonDNApre}
V. Baldazzi, S. Bradde, S. Cocco, E. Marinari, and R. Monasson, Phys. Rev. E 75, 011904 (2007).


\bibitem{monassonsinai2008}
S. Cocco and R. Monasson, Euro. Phys. Lett, 81 20002 (2008).


\bibitem{monassonsinai2009}
S. Cocco and R. Monasson,
Journal of Physics: Conference Series 197, 012005 (2009).






\bibitem{noise}
F. Boettcher, J. Peinke, D. Kleinhans, R. Friedrich, P. G. Lind, M. Haase,
Phys. Rev. Lett. 97, 090603 (2006).


\bibitem{mapsforce}
J.B. Masson, D. Casanova, S. Turkcan, G. Voisinne, M. Popoff, M. Vergassola and A. Alexandrou,
Phys. Rev. Lett 102, 048103 (2009).


\bibitem{biomolecule}
G. Voisinne, A. Alexandrou and J. B. Masson,
Biophysical Journal 98, 596 (2010).

\bibitem{singlemolecule}
S. Turkcan, A. Alexandrou and J.B. Masson,
Biophysical Journal 102, 2288 (2012).

\bibitem{bacterial}
S. Turkcan, J. B. Masson, D. Casanova, G. Mialon, T. Gacoin, J. P. Boilot, M. R. Popoff and A. Alexandrou,
Biophysical Journal 102, 2299 (2012).

\bibitem{decision}
S. Turkcan and J. B. Masson, PLOS ONE 8(12): e82799 (2013).

\bibitem{opticaltweezer}
M. U. Richly, S. Turkcan, A. Le Gall, N. Fiszman, J. B. Masson, N. Westbrook, K. Perronet and A. Alexandrou,
Optics Express 21, 31578 (2013).

\bibitem{membrane}
S. Turkcan S, M. U. Richly, A. Alexandrou, J. B. Masson , PLOS ONE 8(1): e53073 (2013).

\bibitem{proteins}
J. B. Masson, P. Dionne, C. Salvatico, M. Renner, C.G. Specht, A. Triller, and M. Dahan, Biophysical Journal  106, 74 (2014).

\bibitem{diffcoef}
C. L. Vestergaard, P. C. Blainey and H. Flyvbjerg,
Phys. Rev E 89, 022726 (2014).

\bibitem{primerbayesian}
M. El Beheiry, S. Turkcan, M. U. Richly, A.Triller, A. Alexandrou, M. Dahan and J. B. Masson,
Biophysical Journal 110, 1209 (2016).

\bibitem{metzler}
S. Thapa, M. A. Lomholt, J. Krog, A. G. Cherstvy and R. Metzler,
Phys. Chem. Chem. Phys. 20, 29018 (2018).

\bibitem{cherstvy}
A. G. Cherstvy, S. Thapa, C. E. Wagner and R. Metzler,
Soft Matter 15(12), 2526 (2019).


\bibitem{movies}
F. S. Gnesotto, G. Gradziuk, P. Ronceray, C. P. Broedersz,
Nature Communications 11,  5378 (2020).

\bibitem{ronceray2020}
A. Frishman and P. Ronceray,
Phys. Rev. X 10, 021009 (2020).

\bibitem{underdamped}
D. B Bruckner, P. Ronceray, C. P Broedersz
Phys. Rev. Lett. 125, 058103 (2020).

\bibitem{secondorder}
F. Ferretti, V. Chardes, T. Mora, A. M. Walczak, I. Giardina,
Phys. Rev. X 10, 031018 (2020).

\bibitem{learning}
G. Munoz-Gil, M. A. Garcia-March, C. Manzo, J. D. Martin-Guerrero and M. Lewenstein, New J. Phys. 22, 013010 (2020).

\bibitem{cellcell}
D. B. Bruckner, N. Arlt, A. Fink, P. Ronceray, J. O. Radler, C. P. Broedersz, arXiv:2008.03978.



\bibitem{flocking}
A. Cavagna, I. Giardina, F. Ginelli, T. Mora, D. Piovani, R. Tavarone, A. M. Walczak,
Phys. Rev. E 89 042707 (2014).

\bibitem{activeswimmwe}
G. Junot, E. Clement, H. Auradou, R. Garcia-Garcia, arXiv:2012.04528.

\bibitem{activematter}
R. Supekar, B. Song, A. Hastewell, A. Mietke, J. Dunkel , arXiv:2101.06568.



\bibitem{derrida-lecture}
B. Derrida, J. Stat. Mech. P07023 (2007).

\bibitem{harris_Schu}
R. J. Harris and G. M. Sch\"utz,
J. Stat. Mech.  P07020 (2007).

\bibitem{searles}
E. M. Sevick, R. Prabhakar, S. R. Williams, D. J. Searles,
Ann. Rev. of Phys. Chem.  Vol 59, 603 (2008). 

\bibitem{harris}
H. Touchette and R. J. Harris, chapter "Large deviation approach to nonequilibrium systems"
of the book "Nonequilibrium Statistical Physics of Small Systems: Fluctuation Relations and Beyond", Wiley 2013.

\bibitem{mft}
L. Bertini, A. De Sole, D. Gabrielli, G. Jona-Lasinio and C. Landim
Rev. Mod. Phys. 87, 593 (2015).

\bibitem{sollich_review}
R. L. Jack, P. Sollich, The European Physical Journal Special Topics  224, 2351 (2015).

\bibitem{lazarescu_companion}
A. Lazarescu, J. Phys. A: Math. Theor. 48 503001 (2015).

\bibitem{lazarescu_generic}
A. Lazarescu, J. Phys. A: Math. Theor. 50 254004 (2017).

\bibitem{jack_review}
R. L. Jack, Eur. Phy. J. B  93, 74 (2020).


\bibitem{fortelle_thesis}
A. de La Fortelle, PhD (2000)
"Contributions to the theory of large deviations and applications" INRIA Rocquencourt.


\bibitem{vivien_thesis}
V. Lecomte, PhD Thesis (2007)
"Thermodynamique des histoires et fluctuations hors d'\'equilibre"
Universit\'e Paris 7.

\bibitem{chetrite_thesis}
R. Ch\'etrite, PhD Thesis 2008 
"Grandes d\'eviations et relations de fluctuation dans certains mod\`eles de syst\`emes
hors d'\'equilibre",  ENS Lyon.

\bibitem{wynants_thesis}
B. Wynants, arXiv:1011.4210, PhD Thesis (2010), "Structures of Nonequilibrium Fluctuations", Catholic University of Leuven.



\bibitem{chetrite_HDR}
R. Ch\'etrite, HDR Thesis (2018)
"P\'er\'egrinations sur les ph\'enom\`enes al\'eatoires dans la nature",
 Laboratoire J. A. Dieudonn\'e, Universit\' e de Nice.




\bibitem{oono}
Y. Oono,
Progress of Theoretical Physics Supplement 99, 165 (1989).

\bibitem{ellis}
R.S. Ellis, Physica D 133, 106 (1999).

\bibitem{review_touchette}
H. Touchette, Phys. Rep. 478, 1 (2009); \\
H. Touchette, Modern Computational Science 11: Lecture Notes from the 3rd International Oldenburg Summer School, BIS-Verlag der Carl von Ossietzky Universitat Oldenburg, 2011.







\bibitem{fortelle_chain}
G. Fayolle and A. de La Fortelle,
Problems of Information Transmission 38, 354 (2002).

\bibitem{c_largedevdisorder}
C. Monthus, Eur. Phys. J. B 92, 149 (2019) in the
topical issue " Recent Advances in the Theory of Disordered Systems"
edited by F. Igloi and H. Rieger.


\bibitem{c_reset}
C. Monthus, J. Stat. Mech. (2021) 033201.



\bibitem{fortelle_jump}
A. de La Fortelle, 
Problems of Information Transmission 37 , 120 (2001).



\bibitem{maes_canonical}
C. Maes and K. Netocny, Europhys. Lett. 82, 30003 (2008).

\bibitem{maes_onandbeyond}
C. Maes, K. Netocny and B. Wynants, Markov Proc. Rel. Fields. 14, 445 (2008).


\bibitem{chetrite_formal}
A. C. Barato and R. Ch\'etrite, J. Stat. Phys. 160, 1154 (2015).

\bibitem{BFG1}
L. Bertini, A. Faggionato and D. Gabrielli, 
Ann. Inst. Henri Poincare Prob. and Stat. 51, 867 (2015).

\bibitem{BFG2}
L. Bertini, A. Faggionato and D. Gabrielli, 
Stoch. Process. Appli. 125, 2786 (2015).


\bibitem{c_ring}
C. Monthus, J. Stat. Mech. (2019) 023206.

\bibitem{c_interactions}
C. Monthus, J. Phys. A: Math. Theor. 52, 135003 (2019).


\bibitem{c_open}
C. Monthus, J. Phys. A: Math. Theor. 52, 025001 (2019).

\bibitem{barato_periodic}
A. C. Barato, R. Ch\'etrite, J. Stat. Mech. (2018) 053207.

\bibitem{chetrite_periodic}
L. Chabane, R. Ch\'etrite, G. Verley, J. Stat. Mech. (2020) 033208.

\bibitem{c_runandtumble}
C. Monthus, arxiv:2103.08885 and arxiv:2104.10392.






\bibitem{maes_diffusion}
C. Maes, K. Netocny and B.  Wynants,
Physica A 387, 2675 (2008).


\bibitem{engel}
J. Hoppenau, D. Nickelsen and A. Engel,
 New J. Phys. 18 083010 (2016).
 
\bibitem{c_lyapunov}
C. Monthus, J. Stat. Mech. (2021) 033303.



\bibitem{haus}
J. W. Haus et K. W. Kehr,
 Phys. Rep. 150, 263 (1987).

\bibitem{jpb_review}
J. P. Bouchaud and A. Georges,
Phys. Rep. 195, 127 (1990).

\bibitem{annphys90}
J. P. Bouchaud, A. Comtet, A. Georges and P. Le Doussal,
Ann. Phys. 201, 285 (1990).

\bibitem{havlin}
D. Ben-Avraham and S. Havlin, 
``Diffusion and reactions in fractals and disordered systems"
 Cambridge University Press (2000) .

\bibitem{c_review}
C. Monthus, Lett. Math. Phys. 78, 207 (2006).



\bibitem{Der_Pom}
B. Derrida and Y. Pomeau, Phys. Rev. Lett. 48 , 627 (1982).



\bibitem{derrida}
B. Derrida, J. Stat. Phys. 31, 433 (1983).


\bibitem{Kesten}
H. Kesten, Acta Math. 131, 208 (1973);
H. Kesten et al. , Compositio Math 30, 145 (1975).

\bibitem{Solomon}
F. Solomon,  Ann. Proba. 1,31 (1975).

\bibitem{sinai}
Y. G. Sinai,  Theo. Prob. and Appl. 27, 256 (1982).

\bibitem{Der_Hil}
B. Derrida and H. Hilhorst, J. Phys. A 16, 2641 (1983).

\bibitem{Cal}
C. de Callan, J. M. Luck, Th. Nieuwenhuizen and D. Petritis, 
 J. Phys. A 18, 501 (1985).
 

 
 \bibitem{strong_review}
F. Igloi and C. Monthus, Phys. Rep. 412, 277 (2005).


\bibitem{c_microcano}
 C. Monthus,  Phys. Rev. B  69, 054431  (2004).
 
 \bibitem{c_watermelon}
C. Monthus,  J. Stat. Mech.  P06036 (2015).

\bibitem{c_mblcayley}
C. Monthus,  J. Stat. Mech. 123304 (2017).

\bibitem{pldjbp}
T. Gauti\'e, J.P. Bouchaud and P. Le Doussal, arXiv:2101.08082.




\bibitem{jpdir}
J. P. Bouchaud, A. Georges and P. Le Doussal,
 J. Physique 48, 1855 (1987).

\bibitem{aslangul}
C. Aslangul, M. Barthelemy, N. Pottier and D. Saint-James,
 J. Stat. Phys. 59, 11 (1990).

\bibitem{comptejpb}
A. Compte and J. P. Bouchaud,
J. Phys. A. 31, 6113 (1998).


\bibitem{directedtrapandsinai}
C. Monthus, Phys. Rev. E 67, 046109 (2003)


\bibitem{vanwijland_trap}
K. Van Duijvendijk, G. Schehr and F. Van Wijland,
Phys. Rev. E 78, 011120 (2008).

\bibitem{c_ruelle}
C. Monthus, arxiv:2102.10834.


\bibitem{jpb_weak}
J.P. Bouchaud, J. Phys. I (France)  2 , 1705 (1992).

\bibitem{dean}
J. P. Bouchaud and D. Dean, J. Phys. I (France)  5, 265  (1995).


\bibitem{jp_pheno}
C. Monthus and J. P.  Bouchaud,
J. Phys. A 29, 3847 (1996).

\bibitem{bertinjp1}
E. M. Bertin and J. P. Bouchaud, 
Phys. Rev. E 67, 026128 (2003).

\bibitem{bertinjp2}
E. M. Bertin and J. P. Bouchaud, 
Phys. Rev. E 67, 065105(R) (2003).

\bibitem{trapsymmetric}
 C. Monthus,  Phys. Rev. E  68, 036114 (2003). 

\bibitem{trapnonlinear}
 C. Monthus,  J. Phys. A  36, 11605 (2003).  


\bibitem{trapreponse}
 C. Monthus, Phys. Rev. E 69, 026103 (2004).

\bibitem{trap_traj}
M. Ueda and S. Sasa,
J. Phys. A: Math. Theor. 50, 125001 (2017).


\end{thebibliography}
\end{document}